\documentclass{lmcs}
\pdfoutput=1

\usepackage{lastpage}
\lmcsdoi{17}{1}{13}
\lmcsheading{}{\pageref{LastPage}}{}{}%
{Jul.~06,~2017}{Feb.~09,~2021}{}

\usepackage[utf8]{inputenc}

\usepackage{ragged2e}
\newcommand{\Ven}{{\rm v}}
\newcommand{\Int}{{\it Int}}

\newcommand{\CVen}{\textrm{V\!}_C}
\newcommand{\DVen}{\textrm{V\!}_D}
\newcommand{\TVen}{\textrm{V\!}_T}

\def \ndres  {\mathbin{\rlap{\raise.05ex\hbox{$-$}}{\lhd}}}

\newcommand{\rplus}{\real_+}

\usepackage{isabelle,isabellesym}
\usepackage{amssymb}
\usepackage[all]{xy}

\DeclareFontFamily{U}{mathx}{\hyphenchar\font45}
\DeclareFontShape{U}{mathx}{m}{n}{
      <5> <6> <7> <8> <9> <10>
      <10.95> <12> <14.4> <17.28> <20.74> <24.88>
      mathx10
      }{}
\DeclareSymbolFont{mathx}{U}{mathx}{m}{n}
\DeclareFontSubstitution{U}{mathx}{m}{n}
\DeclareMathAccent{\widecheck}{0}{mathx}{"71}
\DeclareMathAccent{\wideparen}{0}{mathx}{"75}

\newcommand{\segset}{S}

\newcommand{\itfin}{{\rm c}}
\newcommand{\itinf}{{\rm so}}

\newcommand{\Sfin}{S_\itfin}
\newcommand{\Sinf}{S_\itinf}

\def \Duration {\textstyle{\int}}

\usepackage{stmaryrd}

\newcommand{\Sup}{\bigsqcup}
\newcommand{\Inf}{\bigsqcap}
\newcommand{\Conv}{\ast}
\newcommand{\sddot}{\ltimes}
\newcommand{\relcomp}{\mathbin{;}}
\newcommand{\relid}{\mathsf{Id}}
\newcommand{\funid}{\mathsf{id}}
\newcommand{\Pow}{\mathcal{P}}

\newcommand{\Supremum}[4]{#1_{#2}^{#3} #4}
\newcommand{\Supr}{\Supremum{\Sup}}
\newcommand{\Infi}{\Supremum{\Inf}}

\newcommand{\A}{{\sf A}}
\newcommand{\B}{{\sf B}}
\newcommand{\C}{{\sf C}}
\newcommand{\D}{{\sf D}}
\newcommand{\E}{{\sf E}}

\newcommand{\LL}{{\sf L}}
\newcommand{\OO}{{\sf O}}
\newcommand{\R}{R}

\newcommand{\T}{{\sf T}}

\newcommand{\calC}{{\mathcal C}}
\newcommand{\calL}{{\mathcal L}}
\newcommand{\calM}{{\mathcal M}}
\newcommand{\calP}{{\mathcal P}}
\newcommand{\calQ}{{\mathcal Q}}
\newcommand{\calR}{{\mathcal R}}
\newcommand{\calS}{{\mathcal S}}
\newcommand{\calT}{{\mathcal T}}
\newcommand{\calX}{{\mathcal X}}
\newcommand{\calY}{{\mathcal Y}}

\newcommand{\converse}[1]{{#1}^{\ensuremath{\smallsmile}}}

\newcommand{\barA}{\converse{\A}}
\newcommand{\barB}{\converse{\B}}
\newcommand{\barD}{\converse{\D}}
\newcommand{\barE}{\converse{\E}}

\newcommand{\barLL}{\converse{\LL}}
\newcommand{\barOO}{\converse{\OO}}

\newcommand{\sem}[1]{#1}

\newcommand{\refprop}[1]{Proposition~\ref{#1}}
\newcommand{\reffig}[1]{Fig.~\ref{#1}}
\newcommand{\refthm}[1]{Theorem~\ref{#1}}
\newcommand{\reflem}[1]{Lem\-ma~\ref{#1}}

\newcommand{\refcor}[1]{Corollary~\ref{#1}}
\newcommand{\refsec}[1]{Section~\ref{#1}}
\newcommand{\refex}[1]{Example~\ref{#1}}

\newcommand{\nat}{\mathbb{N}}
\newcommand{\real}{\mathbb{R}}
\newcommand{\bool}{\mathbb{B}}
\newcommand{\pset}{\mathbb{P}}

\newcommand{\LAB}{\big \langle}
\newcommand{\RAB}{\big \rangle}
\newcommand{\LSB}{\big [}
\newcommand{\RSB}{\big ]}
\newcommand{\VL}{\big |}

\newcommand{\FDia}[1]{\mathop{\VL #1 \RAB}}
\newcommand{\BDia}[1]{\mathop{\LAB #1\, \VL}}
\newcommand{\FBox}[1]{\mathop{\VL #1 \,\RSB}}
\newcommand{\BBox}[1]{\mathop{\LSB#1\, \VL}}

\newcommand{\OMIT}[1]{}

\newcommand{\ch}{\mathbin{\fatsemi}}
\usepackage[colorinlistoftodos]{todonotes}

\definecolor{red}{rgb}{1,0,0}

\definecolor{bdcolor}{rgb}{1,0.5,0.5}
\definecolor{ihcolor}{cmyk}{1,0,1,0}
\definecolor{gscolor}{cmyk}{1,0,0,0}

\def \imp {\Rightarrow}

\def \iff {\Leftrightarrow}

\isabellestyle{it}

\begin{document}

\newcommand{\draftonly}[1]{#1}



\title[Convolution Algebras]{Convolution Algebras: Relational Convolution, Generalised Modalities and Incidence
  Algebras}




\author[B. Dongol]{Brijesh Dongol\rsuper{a}}	
\address{\lsuper{a}University of Surrey, UK}	
\email{b.dongol@surrey.ac.uk}  
\thanks{
The research reported here was supported in part by Australian
Research Council (ARC) Grant DP190102142. Dongol is supported by EPSRC
grants EP/R019045/2, EP/R032556/1 and EP/R025134/2. 
}

\author[I. J. Hayes]{Ian J. Hayes\rsuper{b}}	
\address{\lsuper{b}The University of Queensland, Brisbane, Australia}	
\email{Ian.Hayes@uq.edu.au
}  

\author[G. Struth]{Georg Struth\rsuper{c}}	
\address{\lsuper{c}University of Sheffield, UK}	
\email{g.struth@sheffield.ac.uk
}  


\renewcommand{\theenumi}{\alph{enumi}}

\begin{abstract}
  Convolution is a ubiquitous operation in mathematics and computing.
  The Kripke semantics for substructural and interval logics motivates
  its study for quantale-valued functions relative to ternary
  relations. The resulting notion of relational convolution leads to
  generalised binary and unary modal operators for qualitative and
  quantitative models, and to more conventional variants, when ternary
  relations arise from identities over partial semigroups.
  Convolution-based semantics for fragments of categorial, linear and
  incidence (segment or interval) logics are provided as qualitative
  applications.  Quantitative examples include algebras of durations
  and mean values in the duration calculus.
\end{abstract}

\maketitle


\section{Introduction}\label{sec:introduction}

Convolution is a ubiquitous operation in mathematics, and
computing. Sch\"utzenberger and Eilenberg's approach to formal
languages, for instance, uses convolution of formal power series
(which are functions from free monoids into semirings) to generalise
the standard language product to the context of weighted
automata~\cite{BerstelReutenauer}.  If $\Sigma^\ast$ is the free
monoid over the alphabet $\Sigma$ and $f,g:\Sigma^\ast \to S$ are
functions mapping words in $\Sigma^\ast$ to values or weights in the
semiring $S$, then the convolution of $f$ and $g$ is defined, for all
$x\in \Sigma^\ast$, as
\begin{equation*}
  (f\Conv g)\, x = \sum^{x=yz}_{y,z\in\Sigma^\ast} f\, y \cdot g\, z.
\end{equation*}
Word $x$ is thus split into all possible prefix/suffix pairs $(y,z)$,
the functions $f$ and $g$ are applied separately to $y$ and $z$,
respectively, the weights of $f\, y$ and $g\, z$ are then multiplied
in $S$ and finally the sum of weights for all possible pairs $(y,z)$
is taken. We study convolution algebras with operations of similar
shape in this article, yet replace $\Sigma^\ast$ and $S$ with
different data. 

Other examples relevant to us come from Rota's famous work on the
foundations of combinatorics~\cite{Rota64}, where convolution is one
of the operations of the incidence algebra of segments in locally
finite posets, and Goguen's classical work on fuzzy logic, which uses
a category with quantale-valued sets as objects and quantale-valued
relations as morphisms that are composed by
convolution~\cite{Goguen67}.  Before that, Heisenberg's matrix
approach to quantum mechanics was originally presented as a
convolution algebra of functions from the groupoid of ordered pairs
into the field of complex numbers~\cite{Heisenberg25}, as pointed out
by Connes~\cite{Connes95}.  The Dirichlet convolution of arithmetic
functions and the convolution operation in group algebras used in the
representation theory of finite groups provide even earlier
examples. Notions of convolution in analysis date back at least to
work by Cauchy, but are less relevant for us.

More recently, further computationally interesting applications of
convolution of functions from partial semigroups or monoids into
quantales have been discussed~\cite{DongolHS16}. Separating
conjunction, for instance, is a convolution on a partial abelian
resource monoid within the assertion quantale of separation logic. The
chop modality, a widely used binary modality in interval temporal
logics, arises as convolution on a partial semigroup of intervals and
yields a similar quantale.

It is well known that the logic of bunched implication---the logical
counterpart of the assertion quantale of separation logic---is a
substructural logic similar to relevance and linear logics, which have
Kripke semantics based on ternary frames.  This raises the question
whether convolution generalises similarly to ternary relations, and
hence to generic semantics for substructural logics.  This
generalisation seems interesting for several reasons. Using functions
instead of predicates with modalities supports quantitative
applications. An emphasis on simple uniform constructions on algebras
and mappings between them leads to equally concise formalisations
in proof assistants, and further to simple generic verification
components for separation or interval logics.

The main contribution of this article lies in an answer to that
question: in a novel approach to relational convolution, the
investigation of the generalised binary and unary modalities that
arise from it, in its specialisation to a previous approach based on
quantale-valued functions from partial semigroups~\cite{DongolHS16},
in further instantiations with a focus on incidence algebras and
interval temporal logics, and, last but not least, in
its formalisation in the Isabelle/HOL proof
assistant~\cite{DongolGHS17}.

More specifically, we generalise the standard Kripke semantics for
unary multimodal operators from predicates to lattice-valued functions
and show how quantale modules~\cite{AbramskyV93} and various kinds of
function transformers arise in this setting
(\refsec{sec:modal-binary-relat}). More general notions of binary
modalities and relational convolution over ternary relations are
introduced next (\refsec{sec:modal-ternary-relat}). These generalise
from predicates to quantale-valued functions. J\'onsson and Tarski's
famous duality between $(n+1)$-ary relational structures and boolean
algebras with $n$-ary operators~\cite{JonssonT51} generalises these
results in the arity of relations and modal operators, but for the
special case of powerset algebras. Yet duality theory is not the
subject of this article; see~\cite{HardingWW18} for subsequent results
in the lattice-valued case.  A correspondence theory for relational
convolution is outlined next (\refsec{sec:rel-semigroups}), with
emphasis on relational conditions inducing monoidal laws in the
convolution algebras of quantale-valued functions.  Using these,
previous lifting results to convolution algebras~\cite{DongolHS16} are
generalised: quantale-valued functions with relational convolution as
composition form quantales in the presence of suitable conditions on
relations (\refthm{P:bmod-lifting}).

Theorem \ref{P:bmod-lifting} specialises to a convolution-based
semantics for variants of the Lambek calculus, and hence for fragments
of other substructural logics including categorical and linear ones
(\refsec{sec:lambek}). It also subsumes previous lifting results based
on partial semigroups and monoids (\refsec{sec:partial-semigroups}),
and specialises to various incidence algebras for segments and
intervals in partial orders with different kinds of compositions
(\refsec{sec:segments-intervals}).  Convolution algebras with
semi-infinite intervals (those without upper bounds) are based on
functions from semidirect products of two partial semigroups---one for
finite behaviours, another one for infinite ones---into quantale
modules (\refsec{sec:algebr-semi-infin1} and
\refsec{sec:algebr-semi-infin2}).  They form quantales
(\refthm{P:weak-quantale-lifting}) in which certain distributivity and
unit laws are weakened. A glossary of the main algebraic structures
used is provided in Appendix~\ref{S:alg-summary}.

After these general mathematical investigations, the second part of
this article is devoted to applications. First, a convolution-based
algebraic semantics for Halpern-Shoham and Venema style temporal
logics~\cite{HS91,Venema91} with unary and binary modalities is
presented, but generalised to incidence algebras of segments over
abstract time domains given by arbitrary posets
(\refsec{sec:halpern-shoham-venema}). Within this framework, different
kinds of segments, with or without point segments and with different
kinds of bounds or compositions, can be included in a uniform and
modular way by setting up different kinds of partial semigroups or
monoids. From that basis, a substantial part of the interval temporal
logic ITL~\cite{Mos2012-LMCS} can be obtained, using a semigroup
construction for stream functions that abstract from the dynamics of
state spaces or program stores (\refsec{sec:interval-logics}), and by
instantiating to a time domain of natural numbers. This semantics
extends seamlessly to the duration calculus~\cite{ZH04}
(\refsec{sec:dc}) and one of its variants, the mean-value
calculus~\cite{PandyaR98} (\refsec{sec:mean-value-calculus}), by
instantiating to a real-time domain.  For these last two logics, we
provide examples that illustrate the relevance of convolution in
quantitative modelling, by showing that the algebras of durations and
mean values over intervals form quantales, too.

A convolution-based semantics of separation logic has been
investigated in a previous article~\cite{DongolGS15}.  It provides
further evidence for the universality of the convolution-based
approach to modal and substructural logics outlined.  For all
applications considered, it suffices to specify an appropriate ternary
relation on the fundamental objects considered. These could be
intervals, resources as in separation logic, linear logic or
biological modelling~\cite{PaunRS09}, threads of concurrent
programs~\cite{HMSW11} or even operators representing measurements on
quantum systems~\cite{FoulisB94}.  Often, these relations arise from
equations over partial semigroups or monoids, and from constructions
over these.  The lifting to convolution algebras is then generic, and
it may yield qualitative assertion algebras corresponding to
substructural or modal logics, or else quantitative algebras, for
instance of weights or probabilities---for separation
  logics, interval temporal logics, duration calculi and beyond. A
series of additional examples can be found in~\cite{DongolHS16}.

The main results of this article have been formalised and verified
with the Isabelle/HOL proof assistant. \refsec{sec:isabelle} contains
a brief overview of this implementation. The complete Isabelle code
and a corresponding proof document can be found
online~\cite{DongolGHS17} in the Archive of Formal Proofs~\cite{AFP}.
Due to this, we show only a few proofs in the article; cross references
between the theorems in the article and Isabelle proofs in the Archive of
Formal Proofs can be found in Appendix~\ref{S:crossref}. A typical
lifting result based on convolution, similar to Theorems
\ref{P:bmod-lifting} and \ref{P:weak-quantale-lifting}, can be found
in~\cite{DongolHS16}. The generalised proofs in this article are
similar and provide little further insight.


\section{Generalised Unary Modalities over Binary Relations}
\label{sec:modal-binary-relat}

This section introduces generalised unary modalities, parametrised by
binary relations, that are defined in terms of lattice-valued
functions.  These are related to Halpern-Shoham style interval
modalities~\cite{Venema90,HS91,Venema91} in
Section~\ref{sec:halpern-shoham-venema}. More general notions of
binary modalities, parametrised by ternary relations, are introduced
and related with unary modalities in
Section~\ref{sec:modal-ternary-relat}. Modalities over binary
relations arise in the context of standard Kripke
frames~\cite{BlackburndRV01}.

According to the standard Kripke semantics, if $R\subseteq X\times X$
is a relation and $P\subseteq X$ a predicate, then $\FDia{\R} P\, x$
holds if and only if $R\, x\, y$ and $P\, y$ hold for some
$y\in X$.\footnote{We freely use set and predicate notation for
  relations, we write either $X\to Y$ or $Y^X$ for types and sets of
  functions.}  Similarly, swapping the order of arguments in $R$,
$\BDia{R} P\, y$ holds if and only if $R\, x\, y$ and $P\, x$ hold for
some $x\in X$.  The forward diamond operator
$\FDia{\underline{\phantom{x}}}$ is thus a \emph{relational preimage}
operation; $\FDia{R}P$ yields the set of pre-states that $R$ relates
to any post-state where $P$ holds. The backward diamond
$\BDia{\underline{\phantom{x}}}$ corresponds to a \emph{relational
  image}; $\BDia{R}P$ yields the set of post-states to which $R$
relates any pre-state where $P$ holds. In a Kripke frame, $R$ is
usually interpreted in terms of accessibility or transitions between
possible worlds. Yet some modal logics, such as interval logics,
require other interpretations.

More generally, we assume that $R\subseteq X\times Y$, $f:X \to \calL$
and $g:Y\to \calL$, where $\calL = (L,\leq)$ is a complete lattice.
We write $\sqcup$ for the join and $\sqcap$ for the meet operation in
$\calL$; we write $0$ for the least and $\top$ for the greatest
element in this lattice. Finally, we write
\[
  \Supr{x \in X}{P\, x}{f\, x} = \bigsqcup \{f\, x\mid x \in X \wedge
  P\, x\}
\]
for the supremum of the set $\{f\, x \mid x \in X\wedge P\,
x\}$. Then, for all $x\in X$ and $y\in Y$,
\begin{equation*}
  \label{eq:9}
  \FDia{\R} g\, x = \Supr{y\in Y}{R\, x\, y}{g\, y}\qquad\text{ and }\qquad
  \BDia{\R} f\, y = \Supr{x\in X}{R\, x\, y}{f\, x}.
\end{equation*}
Forward and backward diamonds are related by opposition duality, which
is modelled by conversion: $\BDia{\R} = \FDia{\converse{\R}}$, where
$\converse{R}\, x\, y \Leftrightarrow R\, y\, x$.  Forward and
backward box modalities can be obtained with infima in place of
suprema:
\begin{equation*}
  \FBox{\R} g \, x = \Infi{y \in Y}{R\, x\, y}{g\, y}
  \qquad\text{ and }\qquad
  \BBox{\R} f \, y = \Infi{x \in X}{R\, x\, y}{f\, x}.
\end{equation*}
Whenever the complete lattice is boolean, boxes and diamonds are
related by De Morgan duality.  Using
$\overline{\varphi} = \lambda x.\ \overline{\varphi\, x}$ then yields
$\FBox{\R} g = \overline{\FDia{\R} \overline{g}}$ and
$\BBox{\R} f = \overline{\BDia{\R} \overline{f}}$.

The standard modalities \cite{BlackburndRV01} can be recovered by
restricting types and using $\calL=\mathbb{B}$, the two-element
lattice of booleans. The generalisation to lattice-valued functions
allows the transition from qualitative to quantitative 
reasoning, using, for instance, the complete lattice of extended reals
or the unit interval with respect to $\min$ and $\max$
(cf. \refex{ex:quantales} and Sections~\ref{sec:dc} and
\ref{sec:mean-value-calculus}).

The following statement shows that generalised modalities satisfy
module-like laws, more precisely the laws of \emph{quantale
  modules}~\cite{AbramskyV93}, which are introduced formally in
Section~\ref{sec:algebr-semi-infin2}.  In this lemma, $R \relcomp S$ denotes
the relational composition of $R$ and $S$ and
$\relid_X= \{ (x,x) \mid x \in X\}$ the identity relation over $X$.
\begin{lem}\label{P:unary-module}
  For an index set $I$ and $i \in I$, let
  $R, R_i \subseteq X \times Y$, $S \subseteq Y \times Z$,
  $g, g_i :Y \to \calL$, and $h:Z \to \calL$. Then 
\begin{align}
  \textstyle\FDia{\bigcup_{i\in I}R_i} g & =  \textstyle\Sup_{i\in I}\FDia{\R_i} g,  \label{unary_module_union} \\
  \textstyle\FDia{\R} (\Sup_{i\in I}g_i) & =  \textstyle\Sup_{i\in I}\FDia{\R} g_i,   \label{unary_module_sup} \\
  \FDia{R \relcomp S} h  & =  \FDia{R}(\FDia{S} h),  \label{unary_module_composition} \\
  \FDia{\relid_X} g & = g.   \label{unary_module_identity}
\end{align}
\end{lem}
The proofs have been formalised with Isabelle.  The diamond operator
$\FDia{\underline{\phantom{x}}}: \pset(X\times Y) \to \calL^Y \to
\calL^X$
corresponds to a generalised (module) action of a binary relation of
type $\pset(X\times Y)$ between the complete join (semi)lattices of
functions $\calL^Y$ and $\calL^X$ obtained by pointwise lifting from
$Y$ and $X$.  Identity (\ref{unary_module_composition}) can be written
as $\FDia{R \relcomp S} = \FDia{R}\circ \FDia{S}$, where
$\FDia{R \relcomp S}: \calL^Z \to \calL^X$,
$\FDia{R}: \calL^Y \to \calL^X$ and $\FDia{S}: \calL^Z \to \calL^Y$.
Identity (\ref{unary_module_identity}) can be written as
$\FDia{\relid_X}= \funid_{\calL^X}$, where $\funid_{\calL^X}$ is the
identity function of type $\calL^X\to \calL^X$.  Thus
$\FDia{\underline{\phantom{x}}}$ is indeed a covariant functor between
the category of relations and the category with lattice-valued
functions as objects and higher-order functions---or function
transformers---between these functions as morphisms.  Identity
(\ref{unary_module_union}) can be written as
$\FDia{\bigcup_{i\in I}R_i} = \Sup_{i\in I}\FDia{\R_i} $, hence
$\FDia{\underline{\phantom{x}}}$ sends unions in the category of
relations of type $\pset(X\times Y)$ to suprema in the complete
semilattice of function transformers.  Finally, by
(\ref{unary_module_sup}), $\FDia{\R}$ is (completely) additive and
hence an operator on the lattice of functions $\calL^Y$ in the sense
of boolean algebras with operators~\cite{JonssonT51}.  This justifies
the status of diamond operators as modalities.

Analogous facts for the other kinds of modalities arise by duality.
The operator $\BDia{\underline{\phantom{x}}}$ is contravariant:
$\BDia{R \relcomp S}=\BDia{S}\circ\BDia{R}$ because
$\converse{(R \relcomp S)} = \converse{S} \relcomp \converse{R}$.
Similarly, identities (\ref{unary_module_union}),
(\ref{unary_module_sup}) and (\ref{unary_module_identity}) in
Lemma~\ref{P:unary-module} are dualised by replacing
$\FDia{\underline{\phantom{x}}}$ by
$\BDia{\underline{\phantom{x}}}$. The operator
$\FBox{\underline{\phantom{x}}}$ acts covariantly on the space of
functions of type $\calL^Y\to \calL^X$ under lattice duality like
$\FDia{\underline{\phantom{x}}}$, that is,
$\FBox{R \relcomp S}=\FBox{R}\circ \FBox{S}$ and
$\FBox{\relid_X}=\funid_{\calL^X}$.  However it maps relational unions
to infima in the space of function transformers, i.e.,
$\FBox{\bigcup_{i\in I} \R_i} = \Inf_{i\in I}\FBox{\R_i} $, and it is
(completely) multiplicative, that is,
$\FBox{\R} (\Inf_{i\in I} g_i) = \Inf_{i\in I}\FBox{\R} g_i$.  Once
again, $\BBox{\underline{\phantom{x}}}$ is contravariant,
$\BBox{R \relcomp S}=\BBox{S}\circ\BBox{R}$, and 
the properties corresponding to (\ref{unary_module_union}),
(\ref{unary_module_sup}) and (\ref{unary_module_identity}) in
Lemma~\ref{P:unary-module} arise from the forward box laws by
replacing $\FBox{\underline{\phantom{x}}}$ by
$\BBox{\underline{\phantom{x}}}$.

A study of these relationships in the context of (Sup-)enriched
categories or quantaloids~\cite{Rosenthal91} seems
worthwhile. Pragmatically, however, the functional programming style
used in this section seems general enough to cover various
applications while simple enough for a straightforward formalisation
in an interactive theorem prover like Isabelle.

The final statements of this section show, in the tradition of boolean
algebras with operators, that generalised unary modalities are related
by adjunctions or Galois connections and conjugations.  This yields
additional theorems for free. All four proofs have been verified with
Isabelle.

\begin{lem}\label{P:unary-mod-galois}
  Let $R\subseteq X\times Y$, $f:X\to\calL$ and $g:Y\to\calL$.  Then
  \begin{equation*}
    \FDia{R} g \le f \Leftrightarrow g\le \BBox{R} f
      \qquad\text{ and }\qquad 
    \BDia{R} f \le g \Leftrightarrow f\le \FBox{R} g.
  \end{equation*}
\end{lem}
\begin{lem}\label{P:unary-mod-conjugation}
    Let $R\subseteq X\times Y$, $f:X\to\calL$ and $g:Y\to\calL$, where
    $\calL$ is a complete boolean algebra.  Then
    \begin{equation*}
      \FDia{R} g\sqcap f = 0  \Leftrightarrow g\sqcap
      \BDia{R} f = 0
      \qquad\text{ and }\qquad      
      \FBox{R} f \sqcup g = \top  \Leftrightarrow f \sqcup \BBox{R} g =
      \top.
    \end{equation*}
\end{lem}


\section{Generalised Binary Modalities over Ternary Relations}
\label{sec:modal-ternary-relat}

Kripke frames based on ternary relations yield semantics for
substructural logics such as relevance logics~\cite{DunnRestall02},
the Lambek calculus~\cite{Lambek58}, categorial logics~\cite{MootR12}
or linear logic~\cite{AllweinD93}; the general theory is once again
due to J\'onsson and Tarski~\cite{JonssonT51}.  We generalise the
approach from Section~\ref{sec:modal-binary-relat} to ternary
relations and binary modalities.  These are closely related to the
concatenation or product of the Lambek calculus, the tensor of linear
logic or the chop operators of interval logics. In particular, they
yield relational convolution operators similar to those that appear
widely across mathematics and computing \cite{DongolHS16}. Binary
modalities require enrichment of the complete lattices of the previous
section by an operation of composition.

\begin{defi}\label{def:quantale}
  A \emph{quantale}~\cite{Conway71,Rosenthal90} is a structure
  $\calQ = (Q,\le,\cdot)$ such that $(Q,\le)$ is a complete lattice,
  $(Q,\cdot)$ is a semigroup with composition operator $\cdot$ and the distributivity axioms
\begin{equation*}
  (\Sup\, X)\cdot y = \Supr{x\in X}{}{x \cdot y}
  \qquad\text{ and }\qquad 
  x \cdot \Sup\, Y =  \Supr{y\in Y}{}{x \cdot y}
\end{equation*}
hold for any $X, Y \subseteq Q$.  
We write $0$ for its least and $\top$ for its greatest element with respect to $\le$.
\begin{itemize}\item
A quantale is \emph{unital} if $(Q,\cdot,1)$ is a monoid with unit $1$. 
\item
A quantale is \emph{distributive} if the underlying lattice is.\item A
distributive quantale is \emph{boolean} if every element $x$ is
complemented, that is, $x\sqcap \overline{x} = 0$ and
$x\sqcup \overline{x} = \top$ hold.
\end{itemize}
\end{defi}

The  annihilation laws $x \cdot 0 = 0 =0 \cdot x$ hold in any
quantale because $\Sup \emptyset = 0$.
Our lifting results from quantales to convolution algebras in the
forthcoming sections preserve distributivity and complementation
properties of the quantale.
For the sake of simplicity, however, we present our results for
quantales only and leave the extensions to distributive or boolean
quantales implicit. They can be found in our Isabelle theories.

We present some well known examples of quantales that are useful for our purposes.

\begin{exa}\label{ex:quantales}\hfill
\begin{enumerate}
\item The Booleans form the boolean unital quantale $(\mathbb{B},\le,\sqcap,1)$ in
  which composition is the lattice meet $\sqcap$.  It allows us to treat
predicates as boolean-valued functions.  

\item The Lawvere quantale $(\mathbb{R}_+^\infty,\ge,+,0)$ consists of
  the extended nonnegative reals $\mathbb{R}_+^\infty$ with reversed
  order on the reals $\ge$, $\bigsqcap$ as supremum, $+$ as
  composition extended by $x+\infty=\infty=\infty+x$, and $0$ as its
  unit.

\item Similarly to (a), $(\mathbb{R}_+^\infty,\ge,\max,0)$ forms a unital
  quantale with $\bigsqcap$ as supremum.

\item The structure $(\mathbb{R}_+^\infty,\le,\cdot,1)$ forms a unital quantale
  with $\bigsqcup$ as supremum.

\item The unit interval $([0,1],\le,\cdot,1)$ forms a unital quantale
  with $\bigsqcup$ as supremum. It is isomorphic to the Lawvere
  quantale via the function $(\lambda x \cdot -\ln x)$.

\item The structures $([0,1],\le,\min,1)$ and
$([0,1],\ge,\max,0)$ form unital quantales.

\end{enumerate}
The quantales in (b)-(f) are distributive, but not boolean. \qed
\end{exa}

\begin{defi}\label{def:relation-convolution}
Let $R\subseteq X\times Y\times Z$ be a ternary relation, and let
$f:Y\to \calQ$ and $g:Z\to \calQ$ be functions into the quantale
$\calQ$.  We define, for all $x\in X$, the generalised binary modality
of \emph{relational convolution}  $\Conv_R$ by
\begin{align*}
  (f\Conv_R g)\, x = \Supr{y\in Y,z\in Z}{R\, x\, y\, z}{f\, y\cdot g\, z}.
\end{align*}
\end{defi}
\noindent
In applications, $R$ is often fixed. It is then convenient to simply
write $f \Conv g$.

Sections~\ref{sec:halpern-shoham-venema} and \ref{sec:interval-logics}
show how relational convolution specialises to chop modalities in
various interval logics.  Its relationship to the (non-associative)
Lambek calculus and similar substructural logics is explained in
Section~\ref{sec:lambek}. Its relationship with more conventional
notions of convolution~\cite{DongolHS16}  is discussed further in
Section~\ref{sec:partial-semigroups}. A similar convolution function
into lattices has been studied independently by Harding, Walker and
Walker~\cite{HardingWW18}.

The following binary counterpart of Lemma~\ref{P:unary-module} has
been verified with Isabelle.
\begin{lem}\label{P:binary-module}
  For an index set $I$ and $i\in I$, let
  $R\subseteq X\times Y\times Z$, $f,f_i:Y\to \calQ$ and
  $g,g_i:Z\to \calQ$. Then, with $S = (\bigcup_{i\in I} R_i)$, 
  \begin{align*}
    \textstyle f\Conv_{S} g & =  \textstyle\Sup_{i\in I} f\Conv_{R_i} g,\\
   \textstyle (\Sup_{i\in I} f_i)\Conv_R g & =  \textstyle\Sup_{i\in I} f_i\Conv_R g,\\
    \textstyle f\Conv_R  (\Sup_{i\in I}\, g_i) & =  \textstyle\Sup_{i\in I} f\Conv_R g_i.
  \end{align*}
  \end{lem}
The 
laws (\ref{unary_module_composition}) and (\ref{unary_module_identity}) 
from Lemma~\ref{P:unary-module} make no
sense in this context. 


Next we relate unary and binary modalities.  Both of the following
statements have been verified with Isabelle.
\begin{lem}\label{P:umod-bmod}
  Let $R\subseteq X\times Y$ and $\calQ$ be a unital quantale. Let $f:Y\to \calQ$
  and let the constant function $c_1:Z\to \calQ$ be defined by $c_1\, z = 1$ for
  all $z\in Z$. Then, with $S = \lambda x, y, z.\ R\, x\, y$ and
  $T = \lambda y, x, z.\ R\, x\, y$,
\begin{equation*}
  \FDia{R} f   =  f \Conv_{S} c_1  
\qquad\text{ and } \qquad
  \BDia{R} f   =  f \Conv_{T} c_1.  
\end{equation*}
\end{lem}
\begin{lem}\label{P:bmod-umod}
Let $R\subseteq X\times Y\times Z$, $f:Y\to \calQ$ and $g:Z\to \calQ$. Then, with $S\, x\, (y,z) = R\, x\, y\, z$,
\begin{equation*}
  f\Conv_R g = \FDia{S} (\lambda (y,z).\ (f\, y \cdot g\, z)).
\end{equation*}
\end{lem}
With $(\cdot) = \lambda (y,z).\ y \cdot z$ one can write
$f\Conv_R g = \FDia{S} ((\cdot) \circ (f,g))$ in functional
programming style.

Convolutions $f\Conv g$  have been introduced in Sch\"utzenberger and
Eilenberg's approach to formal language
theory~\cite{BerstelReutenauer}. In this case, $x$, $y$ and $z$ are
words from some free monoid $X^\ast$ and $R\, x\, y\, z$ corresponds
to $x=y\cdot z$. Similar forms of convolution have been considered widely in
mathematics. 
Many well-known constructions from computer science can be represented
this way~\cite{DongolHS16}. 


\section{Relational Semigroups and Convolution Algebras}
\label{sec:rel-semigroups}

We now fix a ternary relation $R\subseteq X\times X\times X$ and write
$f\Conv g$ instead of $f\Conv_R g$. The main result of this section
(Theorem~\ref{P:bmod-lifting}) is also one of the main theorems of
this article. It characterises the convolution algebras that arise
from lifting to function spaces $\calQ^X$ of quantale-valued functions
of type $X\to\calQ$, with relational convolution as the operation of
composition on function spaces and other operations such as suprema
and the order relation lifted pointwise from $Q$. Convolution algebras
are usually quantale-like, but we also encounter situations where
they form semirings.  But before that, in
the tradition of modal correspondence theory, we impose conditions on
$R$ that are reflected by algebraic laws in convolution algebras.

We first consider associativity
$(f \Conv g)\Conv h= f\Conv (g \Conv h)$ of convolution for
quantale-valued functions $f,g,h:X\to \calQ$.  We present two
counterexamples. The first one is computationally interesting, the
second one purely syntactic.

\begin{lem}\label{P:assoccounter}
  There exists a ternary relation over $X$ such that
  $(f\Conv f)\Conv f\neq f\Conv (f \Conv f)$ for some
  $f:X\to \mathbb{B}$.
\end{lem}
\begin{proof}\hfill
\begin{enumerate}
\item\label{P:assoccounter-tree}
  Let $X$ be the set of binary trees with leaves labelled by $a$.
  Let $R\, x\, y\, z$ hold if $y$ is an immediate left subtree of $x$ and $z$
  an immediate right subtree of tree $x$. Let $f=\lambda v.\ (v=a)$.
  Then $f\Conv (f\Conv f)$ holds of the tree below,   whereas $(f\Conv f)\Conv f$ does not.
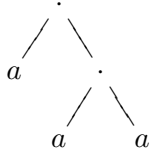
\begin{equation*}
  \def\labelstyle{\normalsize}
\xymatrix @R=1.2pc @C=0.4pc{
& \cdot \ar@{-}[dl]\ar@{-}[dr]\\
a & &\cdot\ar@{-}[dl]\ar@{-}[dr]\\
& a & &a 
}
\end{equation*}
\item\label{P:assoccounter-rel} Let $X=\{a,b\}$,
  $R=\{(a,b,b),(b,b,a)\}$ and consider $f:X\to\mathbb{B}$ defined by
  $f\, a=0$ and $f\, b= 1$.  Then
  $((f\Conv f)\Conv f)\, b = 0 \neq 1 = (f\Conv (f\Conv f))\, b$ holds
  by unfolding definitions and performing some simple calculations. \qedhere
\end{enumerate}
\end{proof}
\noindent
In order to force associativity of convolution, we impose the
following condition on $R$.

\begin{defi}\label{def:relational-semigroup}
  A \emph{relational semigroup} is a structure $\calX = (X,R)$ such
  that $X$ is a set and $R$ a ternary relation over $X$ that satisfies
  the \emph{relational associativity law} for any
  $x, u, v, w \in X$:
\begin{equation*} 
  (\exists y\in X.\ R\, y\, u\, v\wedge R\, x\, y\, w) \Leftrightarrow 
  (\exists z\in X.\ R\,  z\, v\, w\wedge R\, x\, u\, z).
\end{equation*}
\end{defi}
The next result has been verified with Isabelle.
\begin{lem}\label{P:bimod-assoc}
  If $\calX$ is a relational semigroup and $\calQ$ is a quantale, then
  for all $f,g,h:\calX\to \calQ$,
\begin{equation*}
  (f\Conv g)\Conv h= f\Conv(g\Conv h).
\end{equation*}
\end{lem}
Next we consider the unit laws
$f \Conv \funid = f = \funid \Conv f$ for a suitable function $\funid$. Again we
provide a counterexample first.

\begin{lem}\label{P:unitcounter}
  There is a relational semigroup $\calX$ for which there is no
  function $g: \calX \to \mathbb{B}$ such that $f\Conv g = f$ and
  $g\Conv f=f$ hold for all $f:\calX\to \mathbb{B}$.
\end{lem}
\begin{proof}
  Consider the closed strict intervals $[i,j]$ with $i<j$ within
  $[0,1]$ (see Section~\ref{sec:segments-intervals} for formal
  definitions) and let $R\, x\, y\, z$ hold 
  if $y,z\subseteq [0,1]$ are intervals such that
  $x=y\cup z$ whenever the maximal
  point in $y$ equals the minimal point in $z$. As point
  intervals of $[i,i]$ have been excluded by strictness $i<j$,
  all strict intervals $x$ and $y$ satisfy
  $\neg R\, x\, x\, y$ and $\neg R\, x\, y\, x$ and therefore it
  cannot be the case that either
  \begin{eqnarray*}
  (f\Conv g)\, x =\Supr{y,z}{R\, x\, y\, z}{f\, y\cdot g\, z} = f\, x
  \hspace*{2em}\mbox{or}\hspace*{2em}
  (g\Conv f)\, x= \Supr{y,z}{R\, x\, y\, z}{g\, y\cdot f\, z} = f\, x
  \end{eqnarray*}
  for any function $g$, because this would require
  $R\, x\, x\, y$ for some $y$ in the first case and $R\, x\, y\, x$
  for some $y$ in the second one in order to ``filter out'' $f\,
  x$. 
\end{proof}

This proof additionally shows that the candidate identity function $g$
would have to yield $1$ on all intervals $y$ satisfying
$R\, x\, x\, y$ or $R\, x\, y\, x$, assuming those existed, and it
would have to yield $0$ on all other intervals. This motivates the
following definitions.

\begin{defi}\label{def:relational-monoid}
A \emph{relational monoid} is a structure $\calY = (Y,R,\xi)$ such that
$(Y,R)$ is a relational semigroup and $\xi\subseteq Y$ such that for
all $y\in Y$,
\begin{gather*}
  \exists e\in \xi.\ R\, y\, e\, y
  \hspace{1em}\mbox{and}\hspace{1em}
  \exists e\in \xi.\ R\, y\, y\, e,
\end{gather*}
and for all $x,y \in Y$ and $e\in \xi$,
\begin{gather*}
  R\, x\, e\, y  \Rightarrow x=y
  \hspace{1em}\mbox{and}\hspace{1em}
  R\, x\, y\, e  \Rightarrow x=y.
 \end{gather*}
\end{defi}
Using the Kronecker delta function $\delta:Y\to Y\to \calQ$ into a unital
quantale $\calQ$ defined by 
\begin{equation*}
  \delta\, x\, y = 
  \begin{cases}
    1, &\text{if } x = y,\\
0, & \text {otherwise}, 
  \end{cases}
\end{equation*}
we can verify the following fact with Isabelle.
\begin{lem}\label{P:bimod-unit}
  Let $\calY$ be a relational monoid and let
  $\calQ$ be a unital quantale.  Then $\funid =\Sup_{e\in\xi} \delta\, e$ is a left
  and right unit of relational convolution in $\calQ^\calY$.
\end{lem}

Relational encodings of partial algebras date back at least to
Skolem~\cite{Skolem20}.  The correspondence between relational
semigroups and monoids and (partial) algebras is explained in
Section~\ref{sec:partial-semigroups}.  In a monoidal context, a
relation $R\, x\, y\, z$ denotes an identity $x=y\cdot z$, and a
relational specification of an algebraic identity is obtained by
flattening the parse trees of algebraic expressions while memoising
subexpressions.  The unit axioms are more general than those of
monoids in that the order of quantifiers is swapped.  This allows
multiple units in a partial algebra, and different left and right
units for each element, for instance, like in (small) categories. The
precise relationship to categories, and equivalent axiomatisations of
relational monoids with unit axioms more similar to those of
arrows-only categories~\cite{MacLane98}, are discussed
in~\cite{CranchDS20b}. Similar axioms have also been used by
Rosenthal~\cite{Rosenthal97} who has proved a special case of the
following lifting result.

\refthm{P:bmod-lifting}(\ref{P:bmod-lifting-proto}) below does not
hold for quantales but does for proto-quantales.
\begin{defi}\label{def:proto-quantale}
A \emph{proto-quantale} is a quantale in which composition need not
be associative.
\end{defi}
Within this theorem, which has once more been verified by Isabelle,
and beyond we refer to algebraic structures $\calQ^\mathcal{A}$ that
arise from the quantale liftings of suitable relational structures
$\mathcal{A}$ as \emph{convolution algebras}.
\begin{thm}\label{P:bmod-lifting}
  In each $\calQ^{\mathcal{A}}$ below, take composition as convolution
  $\Conv_R$ and suprema and infima to be the pointwise liftings of
  those in $\calQ$.
\begin{enumerate}
\item\label{P:bmod-lifting-proto} 
  Let $X$ be a set and $R \subseteq X \times X \times X$. If $\calQ$ is a
  proto-quantale, then so is $\calQ^{X}$.
\item\label{P:bmod-lifting-quantale}
  Let $\calX$ be a relational semigroup. If $\calQ$ is a quantale, then
  so is $\calQ^\calX$.
\item\label{P:bmod-lifting-unital} Let $\calY$ be a relational
  monoid. If $\calQ$ is a unital quantale, then so is $\calQ^\calY$ with unit $\funid$.
\end{enumerate}
\end{thm}
\noindent
These lifting results extend to distributive and boolean
quantales. Proofs can be found in our Isabelle theories.

Finally, it is natural to consider the law
$\forall x, y, z.\ R\, x\, y\, z \Rightarrow R\, x\, z\, y$, although
it plays no further role in this study.  A relational semigroup is
\emph{abelian} if this \emph{relational commutativity law} holds; we
call a quantale \emph{abelian} if the underlying semigroup is.  We have
verified with Isabelle that, if $\calS$ is an abelian relational
semigroup, then $\calQ^\calS$ is an abelian quantale whenever $\calQ$
is.  In particular, the convolution of quantale-valued
functions from an abelian relational semigroup is commutative. In
addition, we have proved lifting results to abelian quantales similar
to those in Theorem~\ref{P:bmod-lifting}.  These are relevant to
separation logic~\cite{DongolGS15}.

Quantales guarantee that infinite sums or suprema exist.  In other
situations, a restriction to finite sums is possible. 

\begin{defi}\label{def:finitely-decomposable}
A relation $R\subseteq X \times Y\times Z$ is \emph{finitely
  decomposable} if for all $x$ there are finitely many $y$ and $z$
such that $R\, x\, y\, z$ holds. A relational semigroup or monoid is
\emph{finitely decomposable} whenever its relation is.
\end{defi}

This notion adapts the notions of the same name for semigroups and
monoids as well as the topological concept of locally finite collections, and
local finiteness of incidence algebras in order
theory~\cite{Rota64}. Theorem~\ref{P:bmod-lifting} then specialises to
semirings.  On the one hand, these are essentially rings without
additive inverses, that is, the additive reducts of semirings are
abelian monoids, but not necessarily abelian groups.  On the other
hand, they can be seen as quantales in which finite suprema are
replaced by (non-idempotent) sums, whereas infinite sums or infima
need not exist.

\begin{cor}\label{P:locfin-lifting}
  Let $\calS$ be a  finitely decomposable semigroup (monoid). If $\calR$ is
  a (unital) semiring, then so is $\calR^\calS$.
\end{cor}
By finite decomposability, all sums in convolutions $(f\Conv g)\, x$ are
finite and can thus be taken over semirings without convergence
issues. Alternatively one could require that $f$ and $g$ have finite
support. For an extension of this lifting result to
  Kleene algebras see~\cite{CranchDS20}.

Finally, the relation $R\subseteq X\times X\times X$ can be recovered
in $\calQ^X$. The following lemma has not been formalised with
Isabelle. We therefore provide a proof.
\begin{lem}\label{P:rel_embedding}
  A relational monoid $\calY$ can be embedded into
  $\calQ^\calY$ for any unital quantale $\calQ$.
\end{lem}
\begin{proof}
  With $\delta$ as defined above, we have
  $ (\delta\, y \Conv \delta\, z)\, x = \Supr{v,w}{R\, x\, v\,
    w}{\delta\, y\, v \cdot \delta\, z\, w}$ and it follows that
  $((\delta\, y\Conv \delta\, z)\, x = \delta\, x\, x) \Leftrightarrow
  R\, x\, y\, z$, which implies
  $((\delta\, y\Conv \delta\, z) = \delta\, x) \Leftrightarrow R\, x\,
  y\, z$.  Hence consider the relation
  $S\subseteq \calQ^\calY \times \calQ^\calY \times \calQ^\calY$
  defined by $S\, f\, g\, h \Leftrightarrow f = g\Conv h$. Then
  $\delta:\calY\to \calQ^\calY$ is the desired (relational) embedding,
  because, by definition,
  $R\, x\, y\, z \Leftrightarrow S\, (\delta\, x)\, (\delta\, y)\,
  (\delta\, z)$.  The function $\delta$ is injective because if
  $\delta\, x\, z = \delta\, y\, z$ holds for all $z$, then $x=y$.
  The embedding extends to relational monoids because $\delta$ maps
  every $e\in \xi$ to the identity of the convolution, $\funid: \calY\to \calQ$.
\end{proof}


\section{Non-Associative Lambek Calculus and Residuation}
\label{sec:lambek}

Non-associative binary modalities are well known from substructural
logics such as the non-associative Lambek calculus~\cite{MootR12},
which forms a precursor to and fragment of more expressive categorical
and linear logics.  In this case, binary modalities are interpreted
over ternary Kripke frames $R\subseteq X\times X\times X$.
These are sometimes presented as multi-operations or
  hyper-operations of type $X\to X\to \Pow\, X$, which are isomorphic
  to ternary relations~\cite{GalmicheL06}.  In any quantale $\calQ$,
two residuation operations, $\backslash$ and $\slash$ can be defined
for all $u,v,w \in \calQ$, by the Galois connections
\begin{equation*}
u\cdot v\le w ~\Leftrightarrow~ v\le u\backslash w \Leftrightarrow~ u\le w\slash v.
\end{equation*}
In addition to a binary product
modality similar to $f \Conv g$, where $f$ and $g$ are predicates,
hence functions of type $X\to\mathbb{B}$, two residual modalities
$f\backslash g$ and $f\slash g$ can be defined. In our setting,
conflating syntax and semantics, these generalise
to
\begin{equation*}
  (f\backslash g)\, x =\Infi{y,z}{R\, z\, y\, x}{f\, y\backslash g\, z}
  \qquad\text{ and }\qquad 
  (g\slash f)\, x =\Infi{y,z}{R\, z\, x\, y}{g\, z\slash f\, y},
\end{equation*}

In the non-associative Lambek calculus, these lift from the
propositional logic to the modal level. We obtain a more general
result for convolution algebras in our Isabelle theories.
\begin{prop}\label{P:res-lifting}
  Let $R\subseteq X\times Y \times Z$ and let $f:X\to \calQ$, $g:Y\to \calQ$
  and $h:Z\to \calQ$ be functions into a quantale $\calQ$.  Then, 
  \begin{equation*}
  f\Conv g\le h\Leftrightarrow g\le f\backslash h \Leftrightarrow f\le
h\slash g.
  \end{equation*}
\end{prop}

In the correspondence theory of the non-associative Lambek calculus,
relational associativity laws have already been
studied~\cite{MootR12}. In fact, these can be split into two
implications and they are reflected at the modal level. Similarly, one
obtains $(f\Conv g)\Conv h\le f\Conv (g\Conv h)$ and its order dual in
the convolution algebra~\cite{CranchDS20}. 

We also recover the expected relationship between the binary
modalities $f\backslash g$ and $f\slash g$ and unary modal box
operators. If the target quantale forms a complete distributive
lattice and multiplication coincides with meet, then $f\backslash g$
and $f\slash g$ correspond to Heyting implications $f\to g$ and
$f\leftarrow g$. The two cases are distinguished only by the order of
arguments in $R$; they coincide if $R$ is relationally
commutative.  If in addition the lattice is complemented and
$c_0:Z\to \calQ$ defined by $c_0\, z = 0$ for all $z\in Z$, then with $S\, x\, y \Leftrightarrow \exists z.\ R\, z\, y\, x$,
\begin{equation*}
  (f\backslash c_0)\, x = \Infi{y,z}{R\, z\, y\, x}{f\, y \to 0} 
  = \Infi{y,z}{R\, z\, y\, x}{\overline{f}\, y \sqcup 0} 
  =\Infi{y}{S\, x\, y}{\overline{f}\, y} 
  = \FBox{S} \overline{f}\, x.
\end{equation*}

The backward box can be obtained from $c_0\slash f$ by conversion
duality. Deeper investigations of convolution algebras with relational
residuations in other substructural logics, in particular linear ones,
are left for future work.

Modal correspondence theory also studies relational properties induced
by modal ones (conversely to the completeness-like properties in
Section~\ref{sec:rel-semigroups}).  To show that associativity of
relational convolution implies relational associativity, for instance,
one can assume that the latter fails and show that this makes the
former fail as well.  To this end it suffices to check that the
relation $R$ in the proof of
Lemma~\ref{P:assoccounter}(\ref{P:assoccounter-rel}) violates the
relational associativity law, which is routine.
Proofs related to
commutativity and units are similar. Full soundness and completeness
proofs for the Lambek calculus with respect to a relational semantics
have been given by MacCaull~\cite{MacCaull98}, see
also~\cite{AndrekaMikulas94}. For our convolution
  algebras, there are two additional correspondences: properties of
  $\calQ^X$ and $\calQ$ lead to properties of $X$; those of $\calQ^X$
  and $X$ lead to properties of $\calQ$, under certain nondegeneracy
  assumptions on elements of $\calQ$ and $X$,
  respectively~\cite{CranchDS20}. These results are not needed for
  this article.


\section{Partial Semigroups as Relational Semigroups}
\label{sec:partial-semigroups}

This section links relational convolution with more conventional
notions, as investigated in ~\cite{DongolHS16}.  Algebraic semantics for
categorical and linear logics are well known
(see~\cite{AllweinD93,Dosen92,AndrekaMikulas94} for early examples).
We generalise in two ways by considering partial algebras and
quantale-valued functions instead of boolean-values ones.  Lifting
results for functions from partial semigroups and monoids into
convolution algebras formed by quantales are not
new~\cite{DongolHS16}.  Thus it remains to explain how partial
algebras correspond to their relational counterparts. All results in this
section have been verified with Isabelle.

\begin{defi}\label{def:partial-semigroup}
  A \emph{partial semigroup} is a structure $\calS = (X,\cdot,D)$ such
  that $X$ is a set, $D\subseteq X \times X$ the domain of composition
  and $\cdot: D\to X$ a partial operation of composition. Composition
  is associative, $x\cdot (y\cdot z)=(x\cdot y)\cdot z$,
  with respect to Kleene equality in the sense that
  if either side of the equation is defined then so is the other and,
  in that case, both sides are equal.  Formally,
\begin{align*}
    D\,  x\,  y\, \wedge\,  D\,  (x \cdot y)\, z & \Leftrightarrow 
  D\,
    y\, z\,  \wedge  D\, x\,  (y \cdot  z), \\ 
  D\, x\, y\,
  \wedge\, D\,  (x \cdot y)\,  z
  & \Rightarrow  
  (x \cdot y) \cdot z = x \cdot (y \cdot z).
\end{align*}
\end{defi}

\begin{defi}\label{def:partial-monoid}
A \emph{partial monoid}  is a structure 
$\calM = (X,\cdot,D,E)$ such that $(X, \cdot, D)$ is a partial semigroup and 
$E\subseteq X$ a set of (generalised) units that satisfy 
\begin{gather*}
    \exists e \in E.\  D\,  e\,  x  \,\wedge\,  e \cdot x = x, \\
\exists e \in E.\  D\,  x\,  e  \,\wedge\,  x \cdot e = x, \\
 e_1,e_2 \in E \,\wedge\,  D\,  e_1\,  e_2\,\,\Rightarrow\,\, e_1 = e_2.
\end{gather*}
\end{defi}
Every monoid $(X,\cdot,1)$ is a partial monoid with $D=X\times X$ and
$E=\{1\}$. Partial monoids are also related to categories. More
specifically, an object-free category~\cite{MacLane98} is a partial
monoid in which $(x\cdot y)\cdot z$ is defined if and only if
$x\cdot y$ and $y\cdot z$ are both defined. More generally, partial
semigroups are relational semigroups that are \emph{functional} in the
sense that for each $y$ and $z$ there is at most one $x$ such that
$R\, x\,y \,z$; see~\cite{CranchDS20} for further information. This is
the case because the result of the associated multi-operation is
either a singleton set or empty.

\begin{lem}\label{P:ps-rs}
Let $R= \lambda x, y, z.\ D\, y\, z \wedge x=y\cdot z$. 
\begin{enumerate}
\item  If $(X,\cdot,D)$ is a partial semigroup, then 
$(X,R)$ is a relational semigroup.
\item If $(X,\cdot,D,E)$ is a partial monoid, then 
$(X,R,E)$ is a relational monoid.
\end{enumerate}
\end{lem}
\noindent
This immediately yields a previous lifting construction to a
convolution algebra for functions from partial semigroups into
quantales~\cite{DongolHS16} as a corollary to
Theorem~\ref{P:bmod-lifting}.

\begin{cor}\label{P:old-lifting}\hfill
\begin{enumerate}
\item Let $\calS$ 
be a partial semigroup. If $\calQ$ is a quantale,
then so is $\calQ^\calS$.
\item Let $\calM$ be a partial monoid. If $\calQ$ is a unital quantale,
  then so is $\calQ^\calM$.
\end{enumerate}
\end{cor}
\noindent
Again, our Isabelle theories show that these results extend to
distributive or boolean quantales.  Relational convolution now
specialises to the more conventional convolution operation. Using
$x=y\cdot z$ to indicate that $y\cdot z$ is both defined and equal to
$x$, we obtain:
\begin{equation*}
  (f \Conv g)\, x = \Supr{y,z}{x=y\cdot z}{f\, y\cdot g\, z}.
\end{equation*}

For the free semigroup $X^+$ or the free monoid $X^\ast$ over a
finite set $X$, the associated relation $\lambda x,y,z.\ x=y\cdot z$
is finitely decomposable.  As in Corollary \ref{P:locfin-lifting}, the
sum in the convolution can then be taken over an arbitrary semiring
$\calR$.  In formal language theory, functions $\calR^{X^+}$ or
$\calR^{X^\ast}$ are known as \emph{formal power
  series}~\cite{BerstelReutenauer}. These are to weighted automata,
what languages are to ordinary finite state machines.  Languages, in
particular, correspond to functions from $X^+$ or $X^\ast$ into the
semiring of booleans.  In this special case, convolution reduces to
language product.

Finally we present three example constructions that are needed in the
sequel. 

\begin{exa}[Ordered Pairs] \label{ex:ord-pairs} If $X$ is a set,
  then $(X\times X,\cdot,D,E)$ is a partial monoid with
  \begin{align*}
x\cdot y &=(\pi_1\, x,\pi_2\, y),\\
    D &=\{(x,y)\in (X\times X)\times (X\times X) \mid \pi_2\, x = \pi_1\,
  y\},\\
E &=\{(x,x)\mid x\in X\},
  \end{align*}
  where $\pi_1$ and $\pi_2$ are the standard projections.  The
  \emph{cartesian fusion product} $(x_1,x_2)\cdot (y_1,y_2)$ thus
  composes two ordered pairs whenever $x_2=y_1$. This algebra on
  ordered pairs is known as \emph{pair groupoid}. It has been used by
  Heisenberg in his original presentation of matrix
  mechanics~\cite{Connes95}. In particular, $\mathbb{B}^{X\times X}$
  is the quantale of binary relations with convolution as relational
  composition, $\mathbb{Q}^{X\times X}$ is the quantale of $Q$-valued
  relations which Goguen introduced to fuzzy
  logic~\cite{Goguen67}. \qed
\end{exa}

\begin{exa}[Partial Monoid Product]
  \label{ex:part-prod-monoid}
  If $(X,\odot,D_\odot,E_\odot)$ and
  $ (Y,\otimes,D_\otimes,E_\otimes)$ are partial monoids, 
  then $ (X \times Y,\cdot,D,E)$ is a partial
  monoid with
  \begin{align*}
    (x_1,y_1)\cdot (x_2,y_2) &= (x_1 \odot x_2, y_1 \otimes y_2),\\
    D &= \{((x_1,y_1),(x_2,y_2))\mid (x_1,x_2)\in D_\odot \wedge (y_2,y_2)\in
        D_\otimes \},\\
    E&=E_\odot \times E_\otimes. &\tag*{\qed}
  \end{align*}
\end{exa}

\begin{exa}[Monoid-Set Product]
  \label{ex:monoid-set-monoid}
  If $(X,\cdot,D,E)$ is a partial monoid and  $Y$ a
  set, then $(X\times Y,\odot,D',E')$ is a partial monoid with
  \begin{align*}
 (x_1,y)\odot (x_2,y) &= (x_1 \cdot x_2, y),\\
D' &= \{((x_1,y),(x_2, y)) \mid (x_1,x_2)\in D \land y \in Y\},\\
E'&=E \times Y. &\tag*{\qed}
  \end{align*}
\end{exa}


\section{Convolution Algebras of Finite Segments and Intervals}
\label{sec:segments-intervals}

Following the general mathematical considerations thus far, we prepare
for applications to interval logics. Our starting point is Rota's
incidence algebras of order theory~\cite{Rota64}, though we do not
restrict our attention to locally finite posets, which are finitely
decomposable. Instead we focus on quantale-valued functions from
partial algebras of segments and intervals, in line with
Section~\ref{sec:partial-semigroups}.  In that sense, incidence
algebras are convolution algebras that arise from lifting
quantale-valued functions from partial semigroups and monoids of
segments and intervals.

Rota attributes the idea of interval functions to Dedekind and
E. T. Bell. As before, the most important facts in this section have
been verified with Isabelle.  We have so far restricted our Isabelle
formalisation to non-strict closed segments and intervals, that is,
segments or intervals of the form $[i,j]$, with point intervals $[i,i]$
included. Formalisations of strict and (semi-)open intervals are
routine and would not yield additional insights.

\begin{defi}\label{def:segment}
A \emph{segment} of a poset $\calP$ is an ordered pair $(i,j)$ on
$\calP$ in which $i\le j$; the segment is \emph{strict} if $i\neq j$.
We write $S(\calP)$ for the set of all segments and $S_s(\calP)$ for
the set of all strict segments over  $\calP$.  We write $[i,j]$
for the segment $(i,j)$ and write $[i,j]_s$ to indicate
strictness.  
\end{defi}

\begin{defi}\label{def:li-poset}
A \emph{li-poset} is a poset $(\calP,\le)$ that satisfies Halpern and
Shoham's \emph{linear interval property}~\cite{HS91}:
\begin{equation*}
  \forall i,j\in \calP.\ i \le j\ \Rightarrow\ \forall k,l\in \calP.\
  (i \le k \le
  j\, \wedge\,  i \le l\le  j\, \Rightarrow\, k \le l\,
  \vee\,  l \le k).
\end{equation*}
\end{defi}
Intuitively, li-posets generalise linear posets in that all intervals
over li-posets are linear. 

\begin{defi}
  A \emph{(strict) interval} is a (strict) segment of a li-poset.
\end{defi}

Example~\ref{ex:ord-pairs} extends immediately to segments.
\begin{lem}\label{P:seg-sg}
Let $\calP$ be a poset. 
  \begin{enumerate}
  \item $(S_s(\calP),\cdot,D)$ forms a partial semigroup of
    ordered pairs.
\item $(S(\calP),\cdot,D,E)$ forms a partial monoid of
  ordered pairs.
  \end{enumerate}
\end{lem}
\noindent
By \emph{segment fusion}, therefore, $[i,j]\cdot[j',k]=[i,k]$ whenever
$j=j'$.  This is exactly cartesian fusion of ordered pairs. Note that
in each pair, the second component must not be smaller than the first
one.

The proof of Lemma~\ref{P:unitcounter}  shows that partial semigroups $S_s(\calP)$ do not have units. It is now straightforward to lift
from $S_s(\calP)$ and $S(\calP)$ to $\calQ^{S_s(\calP)}$ and
$\calQ^{S(\calP)}$ by virtue of Corollary~\ref{P:old-lifting}.
\begin{cor}\label{P:seg-lifting}
Let $\calP$ be a poset.
\begin{enumerate}
\item If $\calQ$ is a quantale, then so is $\calQ^{S_s(\calP)}$.
\item If $\calQ$ is a unital quantale, then so is $\calQ^{S(\calP)}$.
\end{enumerate}
\end{cor}
\noindent
Relational convolution can now be written, as usual in texts on
incidence algebras, as
\begin{equation*}
  (f \Conv g)\, [i,j] =\Supr{k}{i\le k\le j}{f\, [i,k] \cdot g\, [k,j]}.
\end{equation*}
Finally, using again a Kronecker delta function, the unit of composition on
the incidence algebra is $\delta\, (\pi_1\, x)\, (\pi_2\, x)$ or, more
simply, $\delta\, i\, j$, if $x = [i,j]$.  In the case of locally
finite posets, and in particular intervals formed over $\mathbb{N}$,
Corollary~\ref{P:seg-lifting} generalises to semirings instead of
quantales, as in Corollary~\ref{P:locfin-lifting}.

Segments and intervals are often defined as sets instead of pairs. For
intervals, the associated partial semigroups or monoids are
isomorphic.  For each segment $[i,j]$, the function
$\sigma\, [i,j] =\{k\in \calP\mid i\le k\le j\}$ is a bijective
morphism from the partial semigroup (monoid) of ordered pairs under
interval fusion onto that of set-based intervals under the partial
composition $x\cup y$, whenever $\mathit{max}\, x=\mathit{min}\, y$.
For general segments, however, only
$\sigma\, (x\cdot y) \supseteq \sigma\, x \cup \sigma\, y$ always
holds, and right-hand sides do not generally form segments. For
example, if $x = [i, j]$ and $y = [j, k]$, one may have:
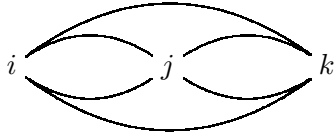
\begin{equation*}
  \def\labelstyle{\normalsize}
  \xymatrix @R=1pc @C=4pc{
    &&\\
    i \ar@/^1pc/@{-}[r]\ar@/_1pc/@{-}[r]\ar@/^2pc/@{-}[rr]\ar@/_2pc/@{-}[rr] & j
    \ar@/^1pc/@{-}[r]\ar@/_1pc/@{-}[r]& k\\
    &&
  }
\end{equation*}
Using segments as sets is therefore not an option. 

Open or semi-open bounded intervals seem less popular in interval
logics.  It then seems appropriate to compose not by fusion, but by
unions provided segments or intervals are adjacent, but
non-overlapping. Such more general segments can be modelled by a
direct product construction like in Example~\ref{ex:part-prod-monoid}.
We outline this construction below; an Isabelle formalisation is left
as future work.

Using the constants $o$ and $c$ to indicate whether
boundaries of segments are open or closed,
an open segment $(i,j)$ can be represented by the pair
$([i,j],(o,o))$, a semi-open segment $(i,j]$ by $([i,j],(o,c))$, a
semi-open segment $[i,j)$ by $([i,j],(c,o))$ and a closed segment
$[i,j]$ by $([i,j],(c,c))$.  

Partial semigroups for strict segments can be constructed from these.
Elements $([i,i],(c,o))$ and $([i,i],(o,c))$ correspond to the empty
segment, which is the unit of composition in the product monoid; for
instance, $[1,3)\cdot [3,3) =[1,3)$ and $[1,3]\cdot (3,3]=[1,3]$.
\begin{lem}\label{P:open-seg-sg}\hfill
\begin{enumerate}
\item Let $C=\{o,c\}$. Then $\calC = (C\times C,\bullet,D_\bullet,E_\bullet)$ is a
  partial monoid of ordered pairs with
  \begin{align*}
    x\bullet y = (\pi_1\, x,\pi_2\, x),\qquad
    D_\bullet =\{(x,y)\mid \pi_2\, x\neq \pi_1\, y\},\qquad
    E_\bullet = \{ (o,c),(c,o) \} 
  \end{align*}
\item $S(\calP)\times \calC$ is a partial product monoid with
  composition, domain of definition and set of units defined as in Example~\ref{ex:part-prod-monoid}.
  \end{enumerate}
\end{lem}
\noindent
Elements $([i,i],(o,o))$, however, shrink segments---$[1,3]\cdot (3,3)=[1,3)$ by definition---and therefore seem undesirable
in the algebra. Fortunately, no element $([i,i],(o,o))$ can be
decomposed into a product of other elements of the product algebra,
and these elements can therefore be neglected.
\begin{lem}\label{P:better-open-seg-sg}
  $(S(\calP)\times \calC) -\{([i,i],(o,o))\mid i\in \calP\}$ forms a
  partial submonoid of $S(\calP)\times \calC$.
\end{lem}
The lifting to  convolution algebras then proceeds as usual by
Corollary~\ref{P:old-lifting}. 


\section{Partial Semigroups of Closed and Semi-Open Segments}
\label{sec:algebr-semi-infin1}

Convolution can be adapted to infinite objects such as infinite words
or streams and, consequently, to semi-infinite intervals $[i,\infty]$
without upper bounds~\cite{DongolHS16}. This extension supports
applications in interval temporal logics and duration calculi. The
following two sections present an alternative to our previous approach
that is based on well known semigroup constructions.

The notion of an \emph{action} of a semigroup or monoid on a set, or
on another semigroup or monoid, is standard.  First we adapt it to
partiality.

\begin{defi}
  A \emph{(left) action} of a partial semigroup
  $\calS = (S,\odot, D_\odot)$ on a partial semigroup
  $\calT = (T,\oplus,D_\oplus)$ is a partial operation
  $\circ: D_\circ \to T$, where
  $D_\circ \subseteq S \times T$, that satisfies
  \begin{align*}
    D_\odot\, s_1\, s_2\, \wedge\,  D_\circ\,  (s_1 \odot s_2)\,  t & \Leftrightarrow
  D_\circ\, s_2\, t\, \wedge\,  D_\circ\,  s_1\,  (s_2 \circ t), \\
  D_\odot\, s_1\, s_2\, \wedge\,  D_\circ\, (s_1 \odot s_2)\, t & \Rightarrow   s_1 \circ (s_2 \circ t) = (s_1 \odot s_2) \circ t,\\
D_\oplus\, t_1\, t_2\, \wedge\,  D_\circ\, s_1\, (t_1 \oplus t_2) & \Leftrightarrow  D_\circ\,
  s_1\, t_1\, \wedge\,  D_\circ\, s_1\, t_2\,  \wedge\, D_\oplus\, (s_1 \circ t_1)\,
  (s_1 \circ t_2), \\ 
  D_\oplus\, t_1\, t_2\, \wedge\, D_\circ\, s_1\,
  (t_1 \oplus t_2) &\Rightarrow  (s_1 \circ t_1) \oplus  (s_1 \circ t_2) = s_1 \circ (t_1 \oplus  t_2).
\end{align*}
If $(S,\odot,D_\odot,E_\odot)$ is a partial monoid, then the left action 
also satisfies the left unit axiom
\begin{equation*}
  e \in E_\odot\ \Rightarrow\ D_\circ\,  e\, t\,  \wedge\,  e \circ t = t.
\end{equation*}
If $(T,\oplus ,D_\oplus,E_\oplus)$ is a partial monoid, then the
following right annihilation axiom also holds:
\begin{equation*}
  e \in E_\oplus\ \Rightarrow\ D_\circ\, s\, e\, \wedge\,  s \circ e = e.
\end{equation*}
\end{defi}

In our intended applications, $\calS$ and $\calT$ represent different
behaviours of a system encoded in terms of pairs
$(s,t)\in \calS\times\calT$. These could be finite or infinite
behaviours, non-faulting or faulting ones or, as in our context, closed
or semi-open intervals or segments, including those with infinite
chains.

More concretely, we consider $\calS$ to be a partial monoid of closed
finite segments and $\calT$ a (partial) monoid of semi-open segments
$[i,j)$. For convenience, we may include segments $[i,\infty)$ and
$[i,\infty]$, which are obtained by adding an element $\infty$ that is
greater than any element of the underlying poset whenever such an
element does not exist. A typical example is formed by the extended
non-negative reals $\mathbb{R}_+^\infty$. General intervals, as in
Section~\ref{sec:segments-intervals}, would require a more tedious
nested product construction.

In this setting, multiplication $\odot$ models fusion of closed
segments, as usual. Action $s\circ t$ represents the fusion of a
closed segment $s$ with a semi-open segment $t$. The use of partial
semigroup actions and the clear typing of closed and semi-open
segments rule out that semi-open segments are fused with closed or
other semi-open ones.

A similar construction on the free monoid $X^\ast$ and the set
$X^\omega$ models the compositions of finite and infinite words,
albeit in a simpler total setting.

Next, in order to construct convolution algebras for such mixed
behaviours, we need to encode their algebras, and in particular those
of closed and semi segments, as partial semigroups.  To this end, we
adapt the well known \emph{semidirect product} construction of two
semigroups or monoids~\cite{CliffordP61}.

\begin{defi}\label{def:semidirect-product}
  For every action of a partial semigroup or partial monoid $\calS$ on
  a partial semigroup $\calT$, the \emph{semidirect product}
  $\calS\sddot \calT=(S\times T,\sddot,D_{\sddot})$ of $\calS$
  and $\calT$ with $\sddot : D_{\sddot}\to S\times T$
  and $D_{\sddot}\subseteq (S\times T)
\times (S\times T)$ is defined by
\begin{align*}
  ((s_1,t_1),(s_2,t_2)) \in D_{\sddot} & \iff D_\odot\, s_1\, s_2\, \wedge\,
  D_\circ\, s_1\, t_2\, \wedge\, D_\oplus\, t_1\, (s_1\circ t_2), \\
  ((s_1,t_1),(s_2,t_2)) \in D_{\sddot} & \Rightarrow 
  (s_1,t_1)\sddot (s_2,t_2) =  (s_1\odot s_2,t_1 \oplus  (s_1\circ t_2)).
\end{align*}
If $\calS$ and $\calT$ are both partial monoids with sets of units $E_\odot$ and
$E_\oplus$, respectively,  then
$\calS\sddot \calT=(S\times T,\sddot ,D_{\sddot},E_{\sddot})$
with $ E_{\sddot} = E_\odot\times E_\oplus$.
\end{defi}

The following fact has been verified with Isabelle.
\begin{prop}\label{P:semigroup-sdprod}
    If $\calS$ and $\calT$ are partial
    semigroups (monoids), then so is $\calS\sddot \calT$.
\end{prop}

Since we are mainly interested in purely closed segments $(s,0)$ and
purely semi-open segments $(0,t)$, we add an element $0$ as an
annihilator to $\calS$ that denotes the empty closed segment. By
definition it satisfies
\begin{equation*} 
0\odot s=0 \qquad \text{ and }\qquad 
0\circ t =0.
\end{equation*}

The use of $\oplus$ in semidirect products requires that we explain
this operation on $\calT$ in the instance of semi-open segments.  In
the context of convolution, where semi-open segments $(0,t)$ are
split with respect to $\sddot$, into all combinations
$(0,t)=(0,t_1\oplus s_1\circ t_2)=(s_1,t_1)\sddot (0,t_2)$, it seems
reasonable to assume that a split produces either $(0,t_1)$ or
$(0,s_1\circ t_2)$. For this we assume that $\oplus$ is not only
associative, but (also commutative and) \emph{selective}:
$t_1\oplus t_2\in \{t_1,t_2\}$ for all $t_1,t_2\in \calT$.  The unit
$0$ in $\calT$ then represents the empty semi-open
segment. Consequently, $s\circ 0 = 0$ holds by definition of partial
monoid actions.

The following example checks that semidirect products of pure closed
and semi-open segments yield the intended behaviour.
\pagebreak[2]
\begin{exa}\label{ex:seg-sdprod}\leavevmode
\begin{enumerate}
\item\label{ex:seg-sdprod-fin-fin}
  $(s_1,0)\sddot (s_2,0)= (s_1\odot s_2,0)$, whenever fusion $s_1
  \odot s_2$
  is defined. Hence, the semidirect product of two closed segments is
  their fusion, as expected.
\item\label{ex:seg-sdprod-inf-inf} $(0,t_1)\sddot (0,t_2)= (0,t_1)$,
  which is always defined, Hence, the fusion of a first semi-open
  segment with a second one simply yields the first segment. This is
  reasonable because one cannot fuse a semi-open segment with any
  other one.
\item\label{ex:seg-sdprod-fin-inf}
  $(s,0)\sddot (0,t)=(0,s\circ t)$, whenever $s\circ t$ is
  defined.  In this case, the closed segment is fused with the semi-open
  segment, as expected.
\item\label{ex:seg-sdprod-inf-fin}
  $(0,t)\sddot (s,0)= (0,t)$, for the same reason as in (\ref{ex:seg-sdprod-inf-inf}). 
\item\label{ex:seg-sdprod-both-both}
  Finally, $(s_1,t_1)\sddot (s_2,t_2)$ equals either
  $(s_1\odot s_2,t_1)$ or $(s_1\odot s_2,s_1\circ t_2)$ by
  selectivity.\qed
\end{enumerate}
\end{exa}

Based on these calculations, it is even simpler to check that
semidirect products of finite and infinite words in $X^\ast$ and
$X^\omega$  model their compositions as expected.


\section{Convolution Algebras of Closed and Semi-Open Segments}
\label{sec:algebr-semi-infin2}

We now construct convolution algebras over partial semigroups of
closed and semi-open segments.  A simplistic approach might attempt
using the partial semigroups and monoids from
Proposition~\ref{P:semigroup-sdprod} together with
Corollary~\ref{P:old-lifting}.  However, this would misrepresent the
most suitable splitting of semi-open segments, intervals or words in
convolutions and therefore the most natural convolution algebra.

\begin{exa}\label{ex:langex}
  Let $f\, x$ state that word $x\in X^\ast\cup X^\omega$ is an element
  of language $f:X^\ast \cup X^\omega\to \mathbb{B}$. Then
  $(f\Conv g)\, x=1$ if and only if $x$ is in the language product of
  $f$ and $g$.  For infinite $x$ this holds if either $f\, x=1$ or $x$
  can be split into some finite $y$ and infinite $z$ such that
  $f\, y=1$ and $g\, z=1$.  This generalises to segments and
  intervals.\qed
\end{exa}

In~\cite{DongolHS16} we have redefined convolution in order to handle
this situation. The convolution algebra then becomes a \emph{weak quantale}.
\begin{defi}\label{def:weak-quantale}
  A quantale is \emph{weak} if the left distributivity law
  $x\cdot \bigsqcup Y= \bigsqcup_{y\in Y} x\cdot y$ holds only for
  $Y\neq\emptyset$ and hence $0$ is no longer a right annihilator.
\end{defi}

\begin{exa}
  In the language algebra $\mathbb{B}^{X^\ast \cup X^\omega}$, 
  products $f\Conv 0$ with the empty language $0$ yield the
  set of all infinite words in $f$ by definition, but not necessarily
  $0$. Once more this generalises to segments and intervals.\qed
\end{exa}

Here, instead of redefining convolution, we adjust the
target algebra $\mathcal{A}$ of functions
$\calS\sddot \calT\to \mathcal{A}$ in such a way that the splitting of
segments according to $\sddot$ is reflected in $\mathcal{A}$. In
addition, it seems reasonable to assume that elements of $\calS$ are
evaluated by a quantalic structure in $\mathcal{A}$ and 
a complete lattice structure $\calT$.
Elements in $\calT$ thus cannot be composed intrinsically by a multiplication.
However, suprema are needed for convolution. 
This leads to the following definition.
\begin{defi}\label{def:quantale-module}
  A \emph{quantale module}~\cite{AbramskyV93} of a quantale
  $\calQ = (Q, \le_\calQ, \cdot)$ and a complete lattice
  $\calL = (L, \le_\calL)$ is an action
  $\circ: \calQ\to \calL\to \calL$ that satisfies, for all $u,v\in Q$,
  $x\in L$, $V \subseteq Q$ and $X \subseteq L$,
\begin{equation*} 
  (u \cdot v)\circ x
  = u \circ (v \circ x),
  \qquad (\Sup V) \circ x 
  = \Supr{v \in V}{}{v \circ x},
  \qquad u \circ \Sup X 
  = \Supr{x \in X}{}{u \circ x}.
\end{equation*}
If $\calQ$ is unital, then, in addition, $1\circ x=x$.
\end{defi}
Obviously, every quantale defines a quantale module on itself with
multiplication as action.  A semidirect product can be defined on
$\calQ$ and $\calL$ as usual as
\begin{equation*}
  (u,x)\sddot (v,y) = (u\cdot v,x \sqcup u \circ y).
\end{equation*}
We can then verify the following counterpart of
Proposition~\ref{P:semigroup-sdprod} with Isabelle.

\begin{prop}\label{P:wquantale-sdprod}
  Let $\calQ$ be a (unital) quantale and $\calL$ a complete
  lattice. Then $\calQ\sddot \calL$ forms a weak (unital) quantale
  in which the left distributivity law is weakened to
  \begin{equation*}
    X\neq \emptyset \Rightarrow u \sddot \Sup\, X = \Supr{x\in X}{}{u\sddot x}
  \end{equation*}
  and where the lattice order and operations are defined as for
  $\calQ\times \calL$.
\end{prop}
The multiplicative unit of $\calQ\sddot\calL$ is $(1,0)$, where $1$ is
the unit of $\calQ$ and $0$ the least element in $\calL$. 
\begin{exa}
  As a counterexample to full left distributivity, hence right
  annihilation, let the unital quantale defined by $0\le \top$ with
  $1=\top$ act on itself. Note that multiplication is fixed. Then,
  $
    (0,\top)\sddot \Sup\emptyset = (0,\top)\sddot (0,0) = (0\cdot 0,1\sqcup
    1\circ 0) = (0,1\sqcup 0)= (0,1)\neq (0,0).
  $
\qed
\end{exa}

Before completing the construction of the convolution algebra we check
that our constructions are consistent with Example~\ref{ex:langex}.

\begin{exa}\label{ex:infinite-fusion}
  Let $\Sfin(\calP)$ and $\Sinf(\calP)$ denote the closed and
  semi-open segments for poset $\calP$ respectively. Assume that
  functions
  $f: \Sfin(\calP) \sddot \Sinf(\calP) \to \calQ\sddot \calL$ are
  given by pairs $f=(f_\itfin,f_\itinf)$ such that
  $f\, (s,t) = (f_\itfin\, s,f_\itinf\, t)$ and that $f_\itfin\, 0 =0$
  and $f_\itinf\, 0=0$. For the sake of simplicity, we further assume
  that $\calQ=\calL$, in which case the action is quantale
  multiplication.
 \begin{itemize}
  \item\label{ex:infinite-fusion-fin} 
    Convolutions over closed segments split only on $\Sfin(\calP)$:
    \begin{align*}
      (f\Conv g)\, (s,0) 
       {} =\Supr{s_1,s_2}{s=s_1\cdot s_2}{(f_\itfin\, s_1,0)\sddot 
        (g_\itfin\ s_2,0)} 
       {} =((f_\itfin\Conv g_\itfin)\, s, 0). 
    \end{align*}
    This recovers the standard convolution of closed segments 
    restricted to $f_\itfin$ and $g_\itfin$. 
    
  \item\label{ex:infinite-fusion-inf} A semi-open segment $(0,y)$ is
    split by convolution into pairs $(x_1,y_1)$ and $(0,y_2)$, and
    hence, $y=y_1\sqcup x_1\circ y_2$. Therefore either $y= y_1$ or
    $y=x_1\circ y_2$ by selectivity and therefore
    \begin{align*}
      (f\Conv g)\, (0,t) 
      & {} = (0,f_\itinf\, t) \sddot (0,0) \sqcup \Supr{s,t'}{t=s\circ t'}{(f_\itfin\, s,0) \sddot (0,g_\itinf\, t')} \\
      & {}= (0,f_\itinf\, t)  \sqcup \Supr{s,t'}{t=s\circ t'}{(0,f_\itfin\, s\cdot g_\itinf\, t')}. 
    \end{align*}
  \end{itemize}
    This is consistent with our previous treatment of convolution~\cite{DongolHS16}.\qed
\end{exa}
The following generalisations of Corollary~\ref{P:old-lifting}, which
we have verified with Isabelle, characterise the convolution algebras
of finite and infinite segments and intervals. The first statement is
generic for partial semigroups and weak quantales.
\begin{thm}\label{P:weak-quantale-lifting}
Let $\calQ$ be a weak quantale.
\begin{enumerate}
\item  If $\calS$ is a partial semigroup, then 
  $\calQ^\calS$ is a weak proto-quantale.
\item  If $\calM$ is a partial monoid, then 
  $\calQ^\calM$ is a weak quantale.
\item If $\calQ$ is unital, then $\funid$ is a left unit
  in $\calQ^\calS$, but not necessarily a right unit.
\end{enumerate}
\end{thm}
\begin{exa} Consider the weak unital quantale defined by
  $0\le 1\le \top$ and by multiplication $0\cdot u=0$,
  $\top\cdot u = \top$.
\begin{enumerate}
\item As a counterexample to associativity, consider this quantale
  with the partial semigroup $\{a,b\}$ where $a^2=b$ is the only
  composition defined, and let $f\ a =f\, b = \top$.  With
  $f^2 = (f \Conv f)$, thus
  $
    (f\Conv f^2)\, b = f\, a \cdot f^2\, a = \top \cdot 0 = \top \neq 0 
    = 0 \cdot \top = f^2\, a \cdot f\, a = (f^2\Conv f)\, b. 
  $
\item As a counterexample to the right unit law, consider this
  quantale with the (total) monoid $\{a,1\}$, where multiplication is
  defined by $a^2=a$.  Let $f\, a=\top$ and $f\, 1=1$.  Then
$
    (f\Conv\funid)\, 1 = f\, 1 \cdot \funid\, 1 \sqcup f\, a \cdot
    \funid\, a = 1 \cdot 1 \sqcup \top \cdot 0 = 1 \sqcup \top = \top
    \neq 1 = f\, 1.
$
\qed
\end{enumerate}
\end{exa}
\noindent
Intuitively, associativity may fail for partial semigroups because not
all elements can be split in such structures. Suprema in convolutions
may thus become empty and associativity fails due to the lack of right
annihilation.  By contrast, the units in partial monoids guarantee
that all elements can be split.

The second statement considers semidirect products, but is still
general.
\begin{prop}\label{P:wq-seg-lifting}
  Let $\calQ$ be a quantale and $\calL$ a complete lattice.
\begin{enumerate}
\item If $\calS$ and $\calT$ are partial semigroups, then   $(\calQ\sddot \calL)^{\calS \sddot \calT}$ is a weak 
  proto-quantale. 
\item If $\calS$ and $\calT$ are partial monoids, then   $(\calQ\sddot \calL)^{\calS \sddot \calT}$ is a weak quantale. 
\item If, in addition, $\calQ$ is unital, then so is
  $(\calQ\sddot \calL)^{\calS \sddot \calT}$.
\item In each case, the subquantale 
  $\calQ^\calS\simeq(\calQ\sddot \{0\})^{\calS \sddot \{0\}}$ is embedded into 
 $(\calQ\sddot \calL)^{\calS \sddot \calT}$. 
\end{enumerate}
\end{prop}
\begin{proof}
  Apart from the right unit law, all properties follow immediately
  from Theorem~\ref{P:weak-quantale-lifting} with
  Propositions~\ref{P:semigroup-sdprod} and~\ref{P:wquantale-sdprod}.  

  The right unit law has not been checked with Isabelle (this would be
  rather tedious), so we provide a proof.  Because the unit in
  $\calQ\times \calL$ is $(1,0)$, the unit $\funid$ on
  $(\calQ\sddot \calL)^{\calS \sddot \calT}$ must map each pair
  $(s,t)$ to $(1,0)$ if $s$ is a monoidal unit segment and $t=0$, and
  to $(0,0)$ otherwise, hence $\funid= (\funid,\lambda x.\ 0)$. We
  calculate
  \begin{align*}
    (f\Conv\funid)\, (s,t) 
& = \Sup_{s_1,s_2,t_1,t_2}^{(s,t)=(s_1,t_1)\sddot (s_2,t_2)}
  (f_\itfin\, s_1,f_\itinf\, t_1) \sddot  (\funid\, s_2,(\lambda x.\
  0)\, t_2)\\
& = \Sup_{s_1,s_2,t_1,t_2}^{s=s_1\odot s_2,t=t_1\oplus s_1\circ t_2}
  (f_\itfin\, s_1 \cdot \funid\,  s_2,f_\itinf\, t_1 \sqcup 
  f_\itfin\,  s_1 \circ 0)\\
    & = \Sup_{s_1,t_1}^{\phantom{\odot s_2\oplus }s=s_1,t=t_1\phantom{s_1\circ t_2}}
  (f_\itfin\, s_1,f_\itinf\, t_1)\\
&= ~f\, (s,t).
  \end{align*}
Finally, it is routine to verify (d) in all cases considered.
\end{proof}

\begin{exa}
The failure of right annihilation with the function
$0=(\lambda x.\ 0,\lambda x.\ 0)$ can be checked by using the
calculation in Example~\ref{ex:infinite-fusion}(b):
\begin{equation*}
  (f\Conv 0)\, (0,t)=(0,f_\itinf\, t) \sqcup \Sup_{s,t'}^{t=s\circ
    t'}(0, f_\itfin
  s\circ (\lambda x.\ 0)\,  t')=(0,f_\itinf\, t)= f\, (0,t)\neq (0,0),
\end{equation*}
whenever $f_\itinf\, t\neq 0$. \qed
\end{exa}

Do Theorem~\ref{P:weak-quantale-lifting}(a) or
Proposition~\ref{P:wq-seg-lifting}(a) rule out associativity of
composition in convolution algebras over partial semigroups of strict
segments and intervals? The pragmatic answer is \emph{no}. By
construction, only splittings of semi-open segments or intervals can
affect associativity (otherwise Corollary~\ref{P:old-lifting} would
already fail). Yet even unbounded intervals $[i,\infty)$ over $\mathbb{N}$ can
always be split finitely many times into finite prefixes and infinite
suffixes. A formalisation of such results with Isabelle is left for
future work.


\section{Modalities over Segments}
\label{sec:halpern-shoham-venema}

The remaining technical sections relate the abstract approach to
segments and their incidence or convolution algebras with well known
interval logics. Binary relationships between intervals were first
proposed by Allen \cite{All83}; a binary modality based on chopping
intervals has been introduced by Moszkowski \cite{MM83}. The modal
logics that arise from such relationships have been studied by Halpern
and Shoham \cite{HS91} (who consider unary modalities only) and
\cite{Venema90,Venema91} (who considers binary modalities including
chop as
well). Decidability, undecidability and completeness of various
fragments including \emph{neighbourhood logic} (see \refsec{sec:dc})
have been studied extensively \cite{ZH04,MM11,MonicaGMS11}.  We refer
to some excellent surveys~\cite{MonicaGMS11,GorankoMS04,Konur} for
more information.

This section outlines how semantics for the interval logics of Halpern
and Shoham~\cite{HS91}, and Venema~\cite{Venema90} arise as instances
of the convolution algebras developed in previous sections. Our
constructions start from partial monoids of strict closed segments
over arbitrary partial orders, yet the approach is modular with
respect to adaptations to partial semigroups of non-strict segments
and to instantiations to algebras of strict or non-strict intervals,
as in Section~\ref{sec:segments-intervals}. It is also generic with
respect to discrete, dense or Dedekind-complete orders.

The unary and binary interval modalities that form our convolution
algebras are more general as well. As in
Sections~\ref{sec:modal-binary-relat} and
\ref{sec:modal-ternary-relat}, they are based on lattice-valued or
quantale-valued functions that admit quantitative interpretations
beyond the standard qualitative ones. As we disregard concrete syntax
in this and the following sections, our algebraic semantics are
\emph{loose}: they are not generated by homomorphic extensions of
semantic maps 
from (finite) sets of atomic functions or predicates and restricted
sets of operations.  Instead they are given by full convolution
algebras formed by all functions or predicates of a certain type. More
precise semantics usually arise as subalgebras induced by homomorphic
images. 
Typical examples are subalgebras
that are closed under the semiring operations, but not with respect to
arbitrary infima and suprema.

We first describe relations for segments analogous to those in Allen's
interval calculus \cite{All83}. These in turn lead to algebraic
companions of Halpern and Shoham's and Venema's interval
logics~\cite{HS91, Venema90} (for segments).  Allen's relations can be
based on a single ternary relation.
\begin{figure}[t]
  \centering 
  \begin{minipage}[t]{0.5\columnwidth}
    \centering 
    \qquad \scalebox{0.8}{\input{graphics/Allen-segment.pspdftex}}
    \caption{$\B$, $\E$, $\A$ and their transposes}
    \label{fig:Allen1}
  \end{minipage}%
  \begin{minipage}[t]{0.5\columnwidth}
    \hfill 
    \scalebox{0.8}{\input{graphics/Allen2-segment.pspdftex}}
    \caption{$\D$, $\OO$, $\LL$ and their 
      transposes}
    \label{fig:Allen2}
  \end{minipage}
\end{figure}

Possible relationships between segments $x$ and $y$ are illustrated in
Figures~\ref{fig:Allen1} and~\ref{fig:Allen2}, recalling that relation
$\converse{R}$ is the converse of $R$. Like Goranko et
al.~\cite{GorankoMS04}, we write $\B$, $\E$ and $\A$ for the
\emph{beginning}, \emph{end} and \emph{after} relationships, and $\D$,
$\OO$ and $\LL$ for the \emph{during}, \emph{overlapping} and
\emph{later} relationships.  Figure~\ref{fig:Allen1} provides an
informal semantics for these. As $\D$, $\OO$ and $\LL$ can be defined
in terms of $\B$, $\E$ and $\A$ (see below), we discuss the latter
three first.

\begin{lem}\label{P:allen-rel}
  Consider the partial semigroup $(\segset(\calP), \cdot, D)$ of
  segments (strict or non-strict) over the poset $\calP$ as a
  relational semigroup $(\calP,\C)$ with
  \begin{equation*}
     \C=\lambda x, y, z.\ D\, y\, z \land x = y \cdot z, 
  \end{equation*}
  like in Lemma~\ref{P:ps-rs}.  Then
\begin{equation*}
  \B\, x\, y \Leftrightarrow \exists z.\ \C\, x\, y\, z,\qquad \E\, x\, y \Leftrightarrow \exists z.\
  \C\, x\, z\, y, \qquad \A\, x\, y \Leftrightarrow \exists z.\ \C\,
  z\, x\, y.
\end{equation*}
\end{lem}
\begin{proof}
  From the definition of $\C$,
  $\B\, x\, y \Leftrightarrow \exists z.\ D\, y\, z \land x = y \cdot
  z$,
  that is, $y$ is indeed a \emph{beginning segment} within $x$.
  Similarly,
  $\E\, x\, y \Leftrightarrow \exists z.\ D\, z\, y \land x = z \cdot
  y$,
  that is, $y$ is indeed an \emph{ending segment} within $x$. Finally,
  $\A\, x\, y \Leftrightarrow \exists z.\ D\, x\, y \land z = x \cdot
  y$,
  that is, segment $y$ comes immediately \emph{after} segment $x$,
  hence can be composed into a segment $z$. More formally, these
  definitions are faithful with respect to those in~\cite{GorankoMS04}
  when specialised to intervals.
\end{proof}

Further, following~\cite{MonicaGMS11,GorankoMS04}, we (re)define
\begin{equation*}
  \D = \B \relcomp \E = \E \relcomp \B,\qquad \OO = \B
  \relcomp\barE,\qquad \LL = \A \relcomp \A.
\end{equation*}

Next we show how the partial and relational semigroups and the
Allen-style segment relations introduced above yield a semantics for
generalised unary Halpern-Shoham modalities over segment functions in
the context of incidence or convolution algebras.  Because forward and
backward modalities are both available, modalities corresponding to
the relations in \reffig{fig:Allen1} can be defined by using only $\B$,
$\E$ and $\A$.

For a segment $x\in S(\calP)$ and quantale-valued function 
$f:S(\calP)\to \calQ$, the intended  semantics is as follows:

\begin{itemize}
\item $\FDia{\B} f\, x$ means that $f$ is applied to some beginning
  segment $y$ of $x$:
\begin{equation*}
  \FDia{\B} f\, x = \Sup^{\B\, x\, y}_{y\in S(\calP)} f\, y =
  \Sup^{\exists z\in S(\calP).\ D\, y\, z \land x=y\cdot z}_{y\in S(\calP)} f\, y;
\end{equation*}
\item $\FDia{\E} f\, x$ means that $f$ is applied to some ending 
segment $y$ of $x$:
\begin{equation*}
  \FDia{\E} f\, x = 
  \Sup^{\exists z\in S(\calP).\ D\, z\, y \land x=z\cdot y}_{y\in S(\calP)} f\, y;
\end{equation*}
\item $\FDia{\A} f\, x$ means that $f$ is applied to some 
segment $y$ that starts precisely where $x$ ends:
\begin{equation*}
  \FDia{\A} f\, x = 
  \Sup^{\exists z\in S(\calP).\ D\, x\, y \land z=x\cdot y}_{y\in S(\calP)} f\, y.
\end{equation*}
\end{itemize}
These equations expand the definitions of unary modalities in
Section~\ref{sec:modal-binary-relat}.

By opposition duality, $\BDia{\B}\, f\, x$ means that $f$ is applied
to some segment of which $x$ is a beginning, $\BDia{E}\, f\, x$ that
$f$ is applied to some segment of which $x$ is an ending, and
$\BDia{\A}\, f\, x$ that $f$ is applied to some segment that ends
precisely where $x$ begins.  The standard interval semantics can be
obtained as before by instantiating $\segset(\calP)$ to li-posets
(Definition~\ref{def:li-poset}) and $\calQ$ to $\mathbb{B}$.  Once
again this is consistent with Section~\ref{sec:modal-binary-relat}.

Next we show how Venema's binary interval modalities $\CVen$, $\DVen$ and
$\TVen$~\cite{Venema91} arise in convolution algebras\footnote{We use
  non-standard notation in order to avoid conflicts within this
  article.}.  These are needed in interval logics because no finite
set of unary interval operators can be functionally complete over over
dense orders~\cite{Venema90}.  The modality $\CVen$, in particular,
corresponds to ITL's chop operator. Like in Lemma~\ref{P:umod-bmod}, 
we then show how Halpern
and Shoham's unary interval modalities can be obtained from Venema's
binary ones by restricting relational convolution.

For predicates $f$ and $g$, the semantics of $\CVen$, $\DVen$ and
$\TVen$ can be described as follows:
  \begin{itemize}
  \item $\CVen\, f\, g\, x=1$ iff there are segments $y$ and $z$ such
    that $x=y\cdot z$ and $f\, y=g\, z=1$;
  \item $\DVen\, f\, g\, x=1$ iff there are segments $y$ and $z$ such
    that $z=y\cdot x$ and $f\, y= g\, z=1$;
  \item $\TVen\, f\, g\, x=1$ iff there are segments $y$ and $z$ such
    that $z=x\cdot y$ and $f\, y= g\, z=1$.
  \end{itemize}
  \begin{figure}[t]
    \centering
    \scalebox{0.8}{\input{graphics/Venema.pspdftex}}
    \caption{$\C$, $\D_\Ven$, $\T_\Ven$}
    \label{fig:Venema}
  \end{figure}
  The standard interval semantics is obtained as before by
  instantiating $\segset(\calP)$ to li-posets and $\calQ$ to
  $\mathbb{B}$.

  Linking this semi-formal semantics with convolution requires ternary
  relations in the context of a partial semigroup
  $(\segset(\calP), \cdot, D)$.  These relations are depicted in
  \reffig{fig:Venema}.  Apart from
  $\C = \lambda x, y, z.\ D\,y\,z \land x=y\cdot z$, they are
  defined as
 \begin{align*}
    \D_\Ven\, x\, y\, z &\, \Leftrightarrow\,  \C\, z\, y\, x \, \Leftrightarrow\,
                          D\, y\, x\wedge z = y \cdot x,\\
    \T_\Ven\, x\, y\, z &\, \Leftrightarrow\,  \C\, z\, x\, y\,
                          \Leftrightarrow\,  D\,
                          x\, y \wedge z= x\cdot y,
  \end{align*}
  and they capture permutations of splittings within and in the
  neighbourhood of a given segment.  A convolution-based semantics for
  generalised Venema modalities is then straightforward.

\begin{lem}\label{P:ven-sem}
  For the partial semigroup $(\segset(\calP), \cdot, D)$ of strict 
  closed segments and functions $f,g: \segset(\calP)\to \calQ$,
  \begin{equation*}
    \CVen\, f\, g = f\Conv_\C g,\qquad
    \DVen\, f\, g  =f\Conv_{\D_\Ven} g,\qquad
    \TVen\, f\, g =
                        f\Conv_{\T_\Ven}  g. 
  \end{equation*}
\end{lem}
\begin{proof}
  Firstly,
  $\CVen\, f\, g = \lambda x.\ \Supr{y,z}{x=y \cdot z}{f\, y \cdot g\,
    z} = f\Conv_\C g$.
  Secondly,
  $\DVen\, f\, g = \lambda x.\ \Supr{y,z}{z=y\cdot x}{f\, y\cdot g\,
    z}=f\Conv_{\D_\Ven} g$.
  Thirdly,
  $\TVen\, f\, g = \lambda x.\ \Supr{y,z}{z=x\cdot y}{f\, y\cdot g\,
    z} = f\Conv_{\T_\Ven} g$.
  In each case, the first step is justified by the semi-formal
  semantics above. Formally, it is easy to show that it is faithful
  with Venema's semantics~\cite{Venema91} when functions are
  specialised to predicates and segments to intervals.
\end{proof}
It can be shown that none of the three modalities can be defined in
terms of permutations and combinations of the other two, and that all
other possible permutations of indices are captured by these three.
It can also be checked that $(\calP,\C)$ forms a relational semigroup,
whereas the structures $(\calP,\DVen)$ and $(\calP,\TVen)$ do not; and hence
the convolutions $\Conv_{\D_\Ven}$ and $\Conv_{\T_\Ven}$ need not be associative.

\begin{lem}\label{P:allen_ven}
  The relations $\C$, $\D_\Ven$ and $\T_\Ven$ are definable
  in terms of Allen's relations (\reffig{fig:Allen1}).
  \begin{align*}
    \C\,x\,y\,z &= 
    \A\, y\, z \land \B\,x\,y \land \E\,x\,z,
    \\
    \D_\Ven\, x\, y\, z &= 
    \A\, y\, x \land \E\, z\, x \land \B\, z\, y,\\
    \T_\Ven\, x\, y\, z  &=  \A\, x\, y \land \B\, z\, x \land \E\,z\, y.
  \end{align*}
\end{lem}

The correspondence between binary and unary modalities captured in
Lemma~\ref{P:umod-bmod} allows us to relate Halpern and Shoham's
modalities with relational convolution.
\begin{lem}\label{P:hs-sem}
 For the partial semigroup $(\segset(\calP), \cdot, D)$ of strict 
  closed segments, $f: \segset(\calP)\to \calQ$ and $c_1=\lambda x.\
  1$, where 1 is the unit in $\calQ$,
\begin{equation*}
  \FDia{\B} f  = f\Conv_\C c_1,\qquad
  \FDia{\E} f = c_1\Conv_\C f,\qquad
  \FDia{\A} f = f \Conv_{\T_\Ven} c_1.
\end{equation*}
\end{lem}
\begin{proof}
  Spelling out the semi-formal semantics above,
  $ \FDia{\B} f\, x = \Supr{y,z}{x = y \cdot z}{f\, y\cdot c_1\, z}$,
  $\FDia{\E} f\, x = \Supr{y,z}{x = y \cdot z}{c_1\, y \cdot f\, z}$
  and
  $\FDia{\A} f\, x = \Supr{y,z}{z = x \cdot y}{f\, y \cdot c_1\, z}$.
\end{proof}
Finally, Lemmas~\ref{P:ven-sem} and \ref{P:hs-sem} in combination
relate Venema's binary segment modalities with Halpern and Shoham's
unary ones:
\begin{equation*}
  \FDia{\B} f = \CVen \, f\, c_1,\qquad \FDia{\E} f = \CVen \, c_1\,
  f,\qquad \FDia{\A} f = \TVen\, f\, c_1.
\end{equation*}


\section{Interval Temporal Logic}
\label{sec:interval-logics}

The generalised segment modalities from
\refsec{sec:halpern-shoham-venema} can be adapted to an algebraic
semantics for the interval temporal logic
ITL~\cite{Mos2012-LMCS,ITLHomepage}. We ignore the next-step operator
in our considerations, and our semantics is once again loose: it is
not generated by a morphism from the ITL syntax. A tighter semantics
would require forming a subalgebra in which arbitrary suprema and
infima need not exist. This would be more akin to a Kleene algebra
without a right annihilator (a.k.a. a weak quantale) than a quantale. Our
loose semantics, however, allows for more expressive higher-order
variants of ITL with quantification over predicates.  We do not
elaborate this in detail and focus on the role of convolution instead.

ITL and the duration calculus, which subsumes it (see
\refsec{sec:dc}), use notions of iteration. These can be defined as
fixpoints on every quantale, including weak ones, due to the
underlying complete lattice structure and monotonicity of the
functions needed.  In particular,
$x\le y \Rightarrow z\cdot x\le z\cdot y$ holds even in weak
quantales. Hence least and greatest fixpoints of
$\varphi=\lambda x.\ f\cdot x$ and $\psi=\lambda x.\ 1\sqcup f\cdot x$
exist and $f^\omega = \nu\varphi$, $f^\ast = \mu\psi$, and
$f^\infty =\nu\psi$ can be used for modelling infinite, finite and
potentially infinite iteration on convolution algebras formed by weak
quantales in our loose semantics.

ITL uses a notion of program store (or state space) that changes over
time within an interval. We model the store dynamics abstractly by
\emph{streams} of type $\calP\to X$ that map a time domain given by a
poset $\calP$ onto a set $X$, which may be a set of functions from
program variables to values.  How the variables and values change over
time, e.g.  by assignment, is not our concern.  In our
  ITL semantics, a predicate $p$ evaluates a stream $\sigma$ over an
  interval $x$, written as $p\, x\, \sigma$. Only the intervals carry
  algebraic structure, $X^\calP$ is merely a set.
The global
relationship between streams, convolution algebras and ITL is captured
as follows.

\newcommand{\FIN}{{\sf Cl}}
\newcommand{\INF}{{\sf Sop}}

\begin{prop}\label{P:itl}
  Let $\FIN = \Sfin(\calP)$ be a partial monoid of (non-strict) closed
  segments under fusion and $\INF = \Sinf(\calP)$ a monoid of
  semi-open segments.  Let $X^\calP$ be a set of streams, let $\calQ$
  be a unital quantale and $\calL$ a complete lattice.
\begin{enumerate}
\item\label{P:itl-fin}  
  $\FIN \times X^\calP$ is a partial monoid and  $\calQ^{\FIN \times
    X^\calP}$ a unital quantale with convolution as composition.
\item\label{P:itl-fin-inf} 
  $(\FIN \sddot \INF)\times X^\calP$ is a partial monoid and
  $(\calQ\sddot \calL)^{(\FIN \sddot\INF) \times X^\calP}$ a
  weak unital quantale, within which the unital quantale $\calQ^{\FIN \times X^\calP}$
  is embedded.
\end{enumerate}
\end{prop}
\begin{proof}
  Part (\ref{P:itl-fin}) is immediate by the product construction in
\refex{ex:monoid-set-monoid} and Corollary~\ref{P:old-lifting}; 
part (\ref{P:itl-fin-inf}) follows from Proposition~\ref{P:semigroup-sdprod} and
Proposition~\ref{P:wq-seg-lifting}. 
\end{proof}
We call the elements of the function spaces \emph{segment stream
  functions}.  For $\calQ=\calL=\mathbb{B}$ we obtain (weak) quantales
of \emph{segment stream predicates} as special cases. These describe
the logic of ITL in terms of convolution algebras.  In the more
concrete case of $\calP=\mathbb{N}$, we may associate $\FIN$ and
$\INF$ with finite and semi-infinite intervals $[i,j]$ and
$[i,\infty)$ in $\mathbb{N}$. Moreover, the order is locally finite
with respect to finite intervals. By Corollary~\ref{P:locfin-lifting},
the convolution algebra of finite intervals then forms an idempotent
semiring (with respect to addition). Similar observations on the
algebra of ITL predicates have been made by H\"ofner and
M\"oller~\cite{HM09,HM08}.

We now explain some of the ITL operations in light of
Proposition~\ref{P:itl} and sketch the most important features of the
loose algebraic semantics.  It is based on interval stream predicates
of type $\Sfin(\nat) \times X^\nat \to \mathbb{B}$ over closed,
non-strict and finite intervals. With this approach, the ITL semantics
of terms or expressions is abstracted into stream functions.

The semantics of boolean operations on predicates is given in
Proposition~\ref{P:itl}(\ref{P:itl-fin}) by the pointwise liftings in
Theorem~\ref{P:bmod-lifting} and Corollary~\ref{P:old-lifting}, owing
to the fact that segment stream functions form partial monoids
(Example~\ref{ex:monoid-set-monoid}).  The semantics of the chop
$p \ch q$ of two predicates $p$ and $q$ is more interesting.
According to the ITL semantics, it holds on some interval $x$ if $x$
can be split into some prefix-suffix pair $y$ and $z$ such that
$x=y\cdot z$, predicate $\sem{p}$ holds on the prefix $y$ and
$\sem{q}$ holds on the suffix~$z$~\cite{Mos2012-LMCS,ITLHomepage}.
Hence chop is indeed convolution and it coincides with $\CVen$. Thus,

\begin{equation*}
  \sem{p \ch q}
  = \lambda x,  \sigma. \Supr{y,z}{x=y\cdot z}{\sem{p}\, y\, \sigma \cdot \sem{q}\, z\, \sigma} 
  = \sem{p}\Conv \sem{q},
\end{equation*}
where the stream $\sigma:\nat\to X$ supplies the store as a function
of time within the intervals $y$ and $z$, over which the predicates
$p$ and $q$ are evaluated.  To obtain the middle expression from
$p\ast q$, the convolution is computed over intervals $x$; the product
and supremum have been extended pointwise with respect to streams
$\sigma$.

The unit predicate is given by $\funid\ [i,j] = \delta\, i\, j$, as in
Section~\ref{sec:segments-intervals}. Intuitively it holds precisely
of any point interval. Finally, the ITL semantics of the iteration
$p^*$, according to which $\sem{p^*}$ holds on interval $x$ if
$\sem{p}$ holds on each interval that results from a finite
decomposition of $x$, can be derived from iteration in the target
quantale.  As the incidence algebra of finite intervals over
$\mathbb{N}$ is locally finite, the idempotent semiring, which forms a
tight ITL semantics as well as a subalgebra of the quantale described
by \refprop{P:itl}(\ref{P:itl-fin}), can be extended to a Kleene
algebra to model iteration of ITL predicates. In sum, the convolution
algebras of a tight convolution-based semantics for ITL predicates
over finite intervals are therefore Kleene algebras.

In the presence of semi-open intervals $\Sinf(\nat)$, the boolean
operations are interpreted as outlined, but chop $p\fatsemi q$ is
interpreted differently, and the incidence algebra is no longer
locally finite.  Now, by the standard ITL semantics, either $\sem{p}$
is evaluated over the entire infinite interval $x$, or the infinite
interval $x$ is split into a finite prefix $y$ and an infinite suffix
$z$, as in Section~\ref{sec:algebr-semi-infin1}, and predicate
$\sem{p}$ is evaluated over $y$, and predicate $\sem{q}$ over $z$.
This corresponds precisely to the treatment of infinite segments in
Example~\ref{ex:infinite-fusion}(\ref{ex:infinite-fusion-inf}).
Hence, also for infinite intervals,
$ \sem{p \ch q} = \sem{p}\Conv \sem{q}$---chop is convolution, as
desired.  In this case, for
Proposition~\ref{P:itl}(\ref{P:itl-fin-inf}), the loose algebraic ITL
semantics is provided by a weak quantale. Tighter semantics in terms
of weak semirings and Kleene algebras, which may arise as subalgebras,
require further work.  Finally, the restriction to finite intervals of
the form $(x,0)$ as in
Example~\ref{ex:infinite-fusion}(\ref{ex:infinite-fusion-fin})
displays the embedding $\lambda x.\ (x,0)$ into the subquantale
$\calQ^{\FIN\sddot X^\calP}$.

    
\section{Duration calculus}
\label{sec:dc}

Duration Calculus (DC) \cite{ZH04,BRC00} is an extension of ITL with
continuous time domain $\mathbb{R}$. This makes DC interesting for
verifying hybrid and cyberphysical systems. In \cite{ZH04}, intervals
are assumed to be finite, non-strict and closed; incidence algebras
are therefore fusion-based. Additionally, DC includes operators for
reasoning about properties in the \emph{neighbourhood} of an interval,
and it offers the capability of measuring and reasoning about
\emph{durations}, that is, the amount of time during which a state
formula holds in an interval \cite{ZH04,HM08,BRC00}. Extensions of DC
admit reasoning with semi-infinite intervals~\cite{ChaochenHX95}. Our
approach supports their uniform treatment via different kinds of
partial semigroups and monoids, as before, and generalisations to
segments.  Hence, we need not make any particular assumptions about
the type of intervals.

Algebraic reconstructions of (fragments of) DC were given previously
by H{\"o}fner and M{\"o}ller \cite{HM08,HM09}, using a
trajectory-based approach. This included embeddings of the
neighbourhood logics into modal semirings~\cite{DesharnaisMS06}, and
the use of weak semirings to cope with infinite intervals. Beyond
that, their approach is unrelated to ours.

We first discuss the duration component, which distinguishes DC from
ITL. As with ITL, we use stream predicates to abstract from the store
dynamics. In DC, these have type $\real\to \bool$, but could easily be
generalised to stream interval predicates of type
$S(\real)\to X^\real \to \bool$, similar to the previous section. We
keep the former for the sake of simplicity.

Intuitively, a duration measures the amount of time for which a
predicate is true in an interval.  Formally, the \emph{duration} of
stream predicate $b$ in interval $x$ is given by
$$
\Duration b\, x = \Duration^{x_{max}}_{x_{min}}~ b\, t\, dt.
$$ 
Hence $\Duration: \bool^\real \to S(\real)\to \rplus^o$, where
$\rplus^o$ for $o\in\{+\infty,-\infty\}$ is an appropriate extension
of the non-negative reals by either $\infty$ or $-\infty$ to indicate
that integrals do not exist, e.g., due to divergence or due to
non-integrable functions.  Note that finitely supported predicates can
be integrable over semi-infinite intervals, and that integrals over
point intervals are zero.
  
Next we outline a convolution-based semantics for predicates, which we
model as interval stream predicates of type
$S(\real) \times \bool^\real \to \bool$.  As in ITL, the meaning of
boolean operators is obtained by pointwise lifting, that of chop
$p \ch q$ is modelled by convolution over finite or semi-infinite
intervals. Beyond that, the semantics of neighbourhood modalities
$\Diamond_r\, p$ (i.e., $p$ holds for some immediately following
interval) and $\Diamond_l\, p$ (i.e., $p$ holds for some immediately
preceding interval) can be obtained from that of the Halpern-Shoham
modalities $\BDia{A}$ and $\FDia{A}$ from
\refsec{sec:halpern-shoham-venema} as $\sem{ \Diamond_r} = \FDia{\A}$
and $\sem{\Diamond_l}\, = \BDia{\A}$.  Since suprema correspond to
existential quantification, this yields
$\sem{\Diamond_r\, p}\, x\, \sigma = \exists y.\ \A\, x\, y \land p \,
y\, \sigma$
and
$\sem{\Diamond_l\, p}\, x\, \sigma = \exists y.\ \A\, y\, x \land p \,
y\, \sigma$.
Finally, the semantics of iteration of predicates again follows ITL.

In light of our mathematical development so far, it is no surprise
that the duration component of DC carries an interesting algebraic
structure too. However, this seems to have been overlooked in the
literature so far. It turns out that DC yields, in fact, an
interesting application of Proposition~\ref{P:itl} and relatives
beyond the booleans in a quantitative setting.

We now study durations as functions from partial monoids of type
$S(\real)$ into the Lawvere quantale $(\rplus^\infty,\ge,+,0)$ from
Example~\ref{ex:quantales}. The following characterisation of the
associated convolution algebra then follows immediately from our
general lifting results, in particular Corollary~\ref{P:old-lifting}.

\begin{prop}\label{P:general-duration-quantale}
  $((\rplus^\infty)^{S(\real)}, \ge, \Conv, \funid)$ is a weak
  distributive unital quantale with 
  \begin{equation*}
    f\Conv g = \lambda x. \bigsqcap_{y,z}^{x = y \cdot z}{f\, y + g\, z}.
  \end{equation*}
\end{prop}
The unit is given, as in Lemma~\ref{P:bimod-unit}, by
$\funid\ x = \bigsqcap_{e\in E} \delta\, e\, x$, where $E$ is the set
of all point intervals.  As always, the delta function yields the unit
of composition of the target quantale when it encounters a point
interval $x$ and the minimal element of the quantale otherwise.  For
the Lawvere quantale these are $0$ and $\infty$, since the order is
reversed.

Proposition~\ref{P:general-duration-quantale} specialises to durations
of predicates $\Duration b$, which have type
$S(\real)\to \rplus^\infty$, as follows, writing
$\Int \subseteq \mathbb{B}^\real$ for the set of all stream
predicates that are integrable over any  strict interval.
\begin{cor}\label{P:duration-quantale}\hfill
  \begin{enumerate}
  \item $\{\Duration b \mid b \in \Int\}$ forms a weak
    distributive quantale.
  \item $\{\Duration b \mid b \in \mathbb{B}^\mathbb{R}\}$ forms a
    weak distributive unital quantale.
  \end{enumerate}
\end{cor}
\noindent
In this instance, we obtain the convolution
\begin{equation*}
  \Duration b\ \Conv \Duration c = \lambda x. \bigsqcap_{y,z}^{x =
    y \cdot z} \Duration b\, y + \Duration c\, z, 
\end{equation*}
and any non-integrable predicate $b$ (for any interval) yields the
same unit $\Duration b$, since $\Duration b\, x$ is equal to $0$ if
$x$ is a point interval and $\infty$ otherwise.

Alternatively, one can use any other of the
$\mathbb{R}^\infty_+$-quantales from Example~\ref{ex:quantales}. These
results develop the quantitative and qualitative aspects of DC
uniformly and algebraically.


\section{Mean-Value Calculus}
\label{sec:mean-value-calculus}

This section briefly discusses the \emph{Mean-Value Calculus} (MVC)
\cite{ZH04,PandyaR98}; an extension of DC that allows reasoning about
the average length of time for which a property holds within an
interval. In our setting, this means that predicates are evaluated in
another quantale, which yields a different quantitative convolution
algebra. 

Now, in addition to the constructs of DC, the \emph{mean value} of an
integrable stream predicate $b \in \Int$ over an interval $x$ is
defined as follows. For the purpose of this section, we assume the
intervals under consideration are finite, however, it is
straightforward to extend the definitions below to the infinite case.
\begin{gather*}
  \theta\, b\, x =
  \begin{cases}
    (\Duration b\, x) / (x_{max} - x_{min} ), & \text{if $x_{max} -
      x_{min} > 0$},\\
    b\, x_{min}, & \text{otherwise}.
  \end{cases}
\end{gather*}
It calculates the proportion of the interval for which $b$ holds as a
value in the unit interval $[0,1]$ in $\real$, hence as a probability, so
that $\theta: \Int
\to S(\real)\to [0,1]$. For a point interval, by definition, the mean
value is the value of $b$ at that point.

To characterise the convolution algebra of mean values of MVC we can
now use the target quantales over $[0,1]$ from
Example~\ref{ex:quantales}, that is, $([0,1],\le,\cdot,1)$,
$([0,1],\le,\min,1)$ or $([0,1],\ge,\max,0)$.
Lifting results similar to Corollary~\ref{P:duration-quantale} are now
straightforward.
\begin{cor}
  \label{lem:mvc-2}
  $\{\theta\, b\mid b\in \Int 
  \}$ forms a weak distributive quantale.
\end{cor}
Depending on the choice of the target quantale, we obtain the following 
convolutions
\begin{align*}
  \theta\, b\, \Conv \, \theta\, c
  &=  \lambda x. \bigsqcup_{y,z}^{x = y \cdot z}{\theta\, b\,y \cdot \theta\,c\,
    z},\\
  \theta\, b\, \Conv\, \theta\, c
  &= \lambda x. \bigsqcup_{y,z}^{x = y \cdot z}{\min\{\theta\,
  b\,y,\theta\, c\, z\}},\\
  \theta\, b\, \Conv\, \theta\, c
  &= \lambda x. \bigsqcap_{y,z}^{x = y \cdot z}{\max\{\theta\, b\,y,\theta\, c\, z\}},
\end{align*}
which are similar to those for durations, but with values taken in
$[0,1]$.


\section{Remarks on the Isabelle Formalisation}\label{sec:isabelle}

Formalising the mathematical structures and theorems in this article
is relatively straightforward using Isabelle, and often leads to
readable definitions and proofs~\cite{DongolGHS17}. Isabelle's
built-in axiomatic type classes allow formalising the basic algebraic
structures used.  Partial semigroups, for instance, extend a
predefined type class that provides an operation of multiplication.
\begin{isabellebody}
\isanewline
\isacommand{class}\isamarkupfalse%
\ partial{\isacharunderscore}semigroup\ {\isacharequal}\ times\ {\isacharplus}\ \isanewline
\ \ \isakeyword{fixes}\ D\ {\isacharcolon}{\isacharcolon}\ {\isachardoublequoteopen}{\isacharprime}a\ {\isasymRightarrow}\ {\isacharprime}a\ {\isasymRightarrow}\ bool{\isachardoublequoteclose}\isanewline
\ \ \isakeyword{assumes}\ mult{\isacharunderscore}assocD{\isacharcolon}\ {\isachardoublequoteopen}D\ y\ z\ {\isasymand}\ D\ x\ {\isacharparenleft}y\ {\isasymcdot}\ z{\isacharparenright}\ {\isasymlongleftrightarrow}\ D\ x\ y\ {\isasymand}\ D\ {\isacharparenleft}x\ {\isasymcdot}\ y{\isacharparenright}\ z{\isachardoublequoteclose}\isanewline
\ \ \isakeyword{assumes}\ mult{\isacharunderscore}assoc{\isacharcolon}\ {\isachardoublequoteopen}D\ x\ y\ {\isasymand}\ D\ {\isacharparenleft}x\ {\isasymcdot}\ y{\isacharparenright}\ z\ {\isasymLongrightarrow}\ {\isacharparenleft}x\ {\isasymcdot}\ y{\isacharparenright}\ {\isasymcdot}\ z\ {\isacharequal}\ x\ {\isasymcdot}\ {\isacharparenleft}y\ {\isasymcdot}\ z{\isacharparenright}{\isachardoublequoteclose}\isanewline
\end{isabellebody}
\noindent Structures that depend on several type parameters, such as
partial actions of partial multiplicative semigroups on sets, can be formalised as locales.
\begin{isabellebody}
\isanewline
  \isacommand{locale}\isamarkupfalse%
\ partial{\isacharunderscore}sg{\isacharunderscore}laction\ {\isacharequal}\ \isanewline
\ \ \isakeyword{fixes}\ Dla\ {\isacharcolon}{\isacharcolon}\ {\isachardoublequoteopen}{\isacharprime}a{\isacharcolon}{\isacharcolon}partial{\isacharunderscore}semigroup\ {\isasymRightarrow}\ {\isacharprime}b\ {\isasymRightarrow}\ bool{\isachardoublequoteclose}\isanewline
\ \ \isakeyword{and}\ act\ {\isacharcolon}{\isacharcolon}\ {\isachardoublequoteopen}{\isacharprime}a{\isacharcolon}{\isacharcolon}partial{\isacharunderscore}semigroup\ {\isasymRightarrow}\ {\isacharprime}b\ {\isasymRightarrow}\ {\isacharprime}b{\isachardoublequoteclose}\ {\isacharparenleft}{\isachardoublequoteopen}{\isasymalpha}{\isachardoublequoteclose}{\isacharparenright}\ \isanewline
\ \ \isakeyword{assumes}\ act{\isacharunderscore}assocD{\isacharcolon}\ {\isachardoublequoteopen}D\ x\ y\ {\isasymand}\ Dla\ {\isacharparenleft}x\ {\isasymodot}\ y{\isacharparenright}\ p\ {\isasymlongleftrightarrow}\ Dla\ y\ p\ {\isasymand}\ Dla\ x\ {\isacharparenleft}{\isasymalpha}\ y\ p{\isacharparenright}{\isachardoublequoteclose}\isanewline
\ \ \isakeyword{and}\ act{\isacharunderscore}assoc{\isacharcolon}\
{\isachardoublequoteopen}D\ x\ y\ {\isasymand}\ Dla\
{\isacharparenleft}x\ {\isasymodot}\ y{\isacharparenright}\ p\
{\isasymLongrightarrow}\ {\isasymalpha}\ {\isacharparenleft}x\
{\isasymodot}\ y{\isacharparenright}\ p\ {\isacharequal}\
{\isasymalpha}\ x\ {\isacharparenleft}{\isasymalpha}\ y\
p{\isacharparenright}{\isachardoublequoteclose}%
\isanewline
\end{isabellebody}
\noindent Note that we write $\alpha\, x\, p$ instead of $x\circ p$,
as $\circ$ is used for function composition in Isabelle. We  extend
this action to an action on a second semigroup:
\begin{isabellebody}
\isanewline
\isacommand{locale}\isamarkupfalse%
\ partial{\isacharunderscore}sg{\isacharunderscore}sg{\isacharunderscore}laction\ {\isacharequal}\ partial{\isacharunderscore}sg{\isacharunderscore}laction\ {\isacharplus}\isanewline
\ \ \isakeyword{assumes}\
act{\isacharunderscore}distribD{\isacharcolon}\
{\isachardoublequoteopen}D\
{\isacharparenleft}p{\isacharcolon}{\isacharcolon}{\isacharprime}b{\isacharcolon}{\isacharcolon}partial{\isacharunderscore}semigroup{\isacharparenright}\
q\ {\isasymand}\ Dla\
{\isacharparenleft}x{\isacharcolon}{\isacharcolon}{\isacharprime}a{\isacharcolon}{\isacharcolon}partial{\isacharunderscore}semigroup{\isacharparenright}\
{\isacharparenleft}p\ {\isasymoplus}\ q{\isacharparenright}\isanewline 
\ \ \ \ {\isasymlongleftrightarrow}\
 Dla\ x\ p\ {\isasymand}\ Dla\ x\ q\ {\isasymand}\ D\ {\isacharparenleft}{\isasymalpha}\ x\ p{\isacharparenright}\ {\isacharparenleft}{\isasymalpha}\ x\ q{\isacharparenright}{\isachardoublequoteclose}\isanewline
\ \ \isakeyword{and}\ act{\isacharunderscore}distrib{\isacharcolon}\ {\isachardoublequoteopen}D\ p\ q\ {\isasymand}\ Dla\ x\ {\isacharparenleft}p\ {\isasymoplus}\ q{\isacharparenright}\ {\isasymLongrightarrow}\ {\isasymalpha}\ x\ {\isacharparenleft}p\ {\isasymoplus}\ q{\isacharparenright}\ {\isacharequal}\ {\isacharparenleft}{\isasymalpha}\ x\ p{\isacharparenright}\ {\isasymoplus}\ {\isacharparenleft}{\isasymalpha}\ x\ q{\isacharparenright}{\isachardoublequoteclose}\ \ \isanewline
\end{isabellebody}
\noindent Proposition~\ref{P:semigroup-sdprod}, which states that the semidirect
product of two partial semigroups forms a partial semigroup, can then
be formalised as a sublocale statement.
\begin{isabellebody}
  \isanewline
\isacommand{definition}\isamarkupfalse%
\ sd{\isacharunderscore}D\ {\isacharcolon}{\isacharcolon}\ {\isachardoublequoteopen}{\isacharparenleft}{\isacharprime}a\ {\isasymtimes}\ {\isacharprime}b{\isacharparenright}\ {\isasymRightarrow}\ {\isacharparenleft}{\isacharprime}a\ {\isasymtimes}\ {\isacharprime}b{\isacharparenright}\ {\isasymRightarrow}\ bool{\isachardoublequoteclose}\ \isakeyword{where}\isanewline
\ \ {\isachardoublequoteopen}sd{\isacharunderscore}D\ x\ y\ {\isasymequiv}\ D\ {\isacharparenleft}fst\ x{\isacharparenright}\ {\isacharparenleft}fst\ y{\isacharparenright}\ {\isasymand}\ Dla\ {\isacharparenleft}fst\ x{\isacharparenright}\ {\isacharparenleft}snd\ y{\isacharparenright}\ {\isasymand}\ D\ {\isacharparenleft}snd\ x{\isacharparenright}\ {\isacharparenleft}{\isasymalpha}\ {\isacharparenleft}fst\ x{\isacharparenright}\ {\isacharparenleft}snd\ y{\isacharparenright}{\isacharparenright}{\isachardoublequoteclose}\isanewline
\isanewline
\isacommand{definition}\isamarkupfalse%
\ sd{\isacharunderscore}prod\ {\isacharcolon}{\isacharcolon}\ {\isachardoublequoteopen}{\isacharparenleft}{\isacharprime}a\ {\isasymtimes}\ {\isacharprime}b{\isacharparenright}\ {\isasymRightarrow}\ {\isacharparenleft}{\isacharprime}a\ {\isasymtimes}\ {\isacharprime}b{\isacharparenright}\ {\isasymRightarrow}\ {\isacharparenleft}{\isacharprime}a\ {\isasymtimes}\ {\isacharprime}b{\isacharparenright}{\isachardoublequoteclose}\ \isakeyword{where}\isanewline
\ \ {\isachardoublequoteopen}sd{\isacharunderscore}prod\ x\ y\ {\isacharequal}\ {\isacharparenleft}{\isacharparenleft}fst\ x{\isacharparenright}\ {\isasymodot}\ {\isacharparenleft}fst\ y{\isacharparenright}{\isacharcomma}\ {\isacharparenleft}snd\ x{\isacharparenright}\ {\isasymoplus}\ {\isacharparenleft}{\isasymalpha}\ {\isacharparenleft}fst\ x{\isacharparenright}\ {\isacharparenleft}snd\ y{\isacharparenright}{\isacharparenright}{\isacharparenright}{\isachardoublequoteclose}\ \ \ \isanewline
\ \ \isanewline
\isacommand{sublocale}\isamarkupfalse%
\ dp{\isacharunderscore}semigroup{\isacharcolon}\ partial{\isacharunderscore}semigroup\ sd{\isacharunderscore}prod\ sd{\isacharunderscore}D\isanewline
\ \ $\langle$proof$\rangle$
\isanewline
\end{isabellebody}
\noindent The sublocale statement requires a signature matching; here
the instances of the product operation and the domain of definition of
the partial semigroup must be declared.  We supply a semidirect
product operation and its domain of definition, which have been
defined before the sublocale statement.

Relational convolution can be defined in
Isabelle as follows.
\begin{isabellebody}
  \isanewline
\isacommand{definition}\isamarkupfalse%
\ bmod{\isacharunderscore}comp\
{\isacharcolon}{\isacharcolon}\isanewline 
\ \  {\isachardoublequoteopen}{\isacharparenleft}{\isacharprime}a\ {\isasymRightarrow}\ {\isacharprime}b\ {\isasymRightarrow}\ {\isacharprime}c\ {\isasymRightarrow}\ bool{\isacharparenright}\ {\isasymRightarrow}\ {\isacharparenleft}{\isacharprime}b\ {\isasymRightarrow}\ {\isacharprime}d\ {\isacharcolon}{\isacharcolon}\ proto{\isacharunderscore}quantale{\isacharparenright}\ {\isasymRightarrow}\ {\isacharparenleft}{\isacharprime}c\ {\isasymRightarrow}\ {\isacharprime}d{\isacharparenright}\ {\isasymRightarrow}\ {\isacharprime}a\ {\isasymRightarrow}\ {\isacharprime}d{\isachardoublequoteclose}\ {\isacharparenleft}{\isachardoublequoteopen}{\isasymotimes}{\isachardoublequoteclose}{\isacharparenright}\ \isakeyword{where}\ \isanewline
\ \ {\isachardoublequoteopen}{\isasymotimes}\ R\ f\ g\ x\ {\isacharequal}\ $\bigsqcup${\isacharbraceleft}f\ y\ {\isasymcdot}\ g\ z\ {\isacharbar}y\ z{\isachardot}\ R\ x\ y\ z{\isacharbraceright}{\isachardoublequoteclose}\isanewline
\isanewline
\isacommand{definition}\isamarkupfalse%
\ {\isachardoublequoteopen}f\ {\isasymstar}\ g\ {\isacharequal}\ {\isasymotimes}\ R\ f\ g{\isachardoublequoteclose}\isanewline 
\end{isabellebody}
\noindent The convolution operation $\otimes$ is supplied with a
ternary relation $R$ (in previous sections we have written $\Conv_R$),
and then with functions $f$, $g$ and an element $x$ of suitable
type. Isabelle allows us to write $f\star g$ (instead of $f\Conv_R g$)
when $R$ is fixed.  In this example, the sort of the output of the
function has been restricted to proto-quantales. In our formalisation
we use even more general kinds of quantales.  Our formalisations of
variants of quantales are once more based on type classes. As a final
example, we show how the lifting result from
Theorem~\ref{P:bmod-lifting}(\ref{P:bmod-lifting-quantale}) can be
captured as an instantiation statement in Isabelle.

\begin{isabellebody}
\isanewline
\isacommand{instantiation}\isamarkupfalse%
\ {\isachardoublequoteopen}fun{\isachardoublequoteclose}\
{\isacharcolon}{\isacharcolon}\
{\isacharparenleft}rel{\isacharunderscore}semigroup{\isacharcomma}quantale{\isacharparenright}\
quantale\isanewline
\ \ $\langle$proof$\rangle$
\isanewline
\end{isabellebody}

\noindent As expected, it states that functions from a relational
semigroup, which has been formalised as a type class, into a quantale
forms an instance of a quantale.  

All constructions of partial semigroups and convolution
algebras in the paper have been formalised by similar sublocale or
instantiation statements, or by interpretation statements that are
similar to instantiations. 

Isabelle offers a range of proof tools to explore the structures in
this article and reason about them.  First of all, its
counterexample generators are helpful, for instance, for debugging
theories. Automated theorem provers, SMT solvers and built-in
simplifiers yield a high degree of proof automation for simple
equational reasoning with first-order structures. Reasoning with
higher-order structures, such as quantales and convolutions, however
may require a significant amount of user interaction and a granularity
of proof much finer than that of mathematical textbooks.

Our entire formalisation can be found online in the Archive of Formal
Proofs~\cite{DongolGHS17}.  So far it covers most of the technical
material up to Section~\ref{sec:halpern-shoham-venema}; open and
semi-closed intervals being a notable exception.  Results on duration
and mean-value calculi have not been included because in particular the
construction of quantales over the extended non-negative reals or the
unit interval require some background theory development for these
number domains to be encoded within Isabelle. All theorems that have
not been formalised are mentioned explicitly in the article; a list of
cross-references between all results in the article and those in the
Archive of Formal Proofs can be found in Appendix~\ref{S:crossref}.

Our Isabelle convolution components can serve as a basis for (a)
formalising the concrete interval logics described in
Section~\ref{sec:halpern-shoham-venema}-\ref{sec:mean-value-calculus}
and (b) building verification components for these using a shallow
embedding of our algebraic semantics.  In addition, our Isabelle
components for relational convolution form a basis for formalised
reasoning about resources, as for instance in separation
logic~\cite{Reynolds02}, and for formalising a wide range of models of
computational interest, from (quantale-valued) relations and
(weighted) languages to program traces, partial-order semantics for
concurrency and even quantum logics in a uniform way, simply by
setting up the appropriate partial semigroups~\cite{DongolHS16}.

Experience shows that the simple axiomatic approach to algebras that
underlies our formalisation is sufficient for many verification
applications~\cite{ArmstrongGS16,DongolGS15,GomesS16}.
An in-depth formalisation of (partial) semigroups, their morphisms and
subalgebras, however, requires the explicit consideration of carrier
sets, for which our current approach is too limited. A categorical
formalisation of the topics investigated may not even be feasible with
Isabelle.


\section{Conclusions}
\label{sec:conclusion}

The main aim of this article has been a generalisation of our previous
approach to convolution as a universal operation in
computing~\cite{DongolHS16} to ternary relations.  While the emphasis
of the applications considered was on (generalised) interval logics,
separation logic, in particular the view of separating conjunction as
convolution, has been considered in a companion
paper~\cite{DongolGS15}.  In all these cases, the general approach
consists in setting up the appropriate ternary relations, which are
often generated by partial semigroups, partial monoids or combinations
of these, and then using the general lifting construction to build
a convolution algebra.  If the target quantale used in the lifting
is formed by the booleans, then the convolution algebra is an algebra
of predicates, hence the lifting embodies a powerset construction with
convolution as complex product. In more general cases, convolution
algebras capture quantative aspects of computing systems such as
durations, weights or probabilities, as our examples show.

The main features of the approach, including the most important
lifting theorems, have already been formalised in
Isabelle~\cite{DongolGHS17}.  The resulting mathematical components
provide first of all a basis for the design of verification
components, which are currently under construction for separation
logic and concurrent Kleene algebras. Similar components for interval
logics and Duration Calculus, with applications in hybrid and
cyberphysical systems verification, could be obtained along the same
lines as instances of the general approach.  Secondly, most of the
computationally interesting models of variants of Kleene algebras
within the Archive of Formal Proofs~\cite{AFP}, which include
relations, languages, sets of paths in a digraph, program traces, and
matrices over Kleene algebras, could be obtained via convolution
simply by setting up the appropriate partial semigroups.

Beyond that we envisage various avenues for future research.  These
include the investigation of other substructural logics, in particular
linear logics, and the effect algebras that arise in the foundations
of quantum mechanics as convolution algebras, the exploration of other
quantitative applications that arise within stochastic or
probabilistic systems, and last but not least,  a consideration of the
approach in the realm of higher category theory.


\section*{Acknowledgments}
  \noindent The authors would like to thank Mark
Hepple for discussions on the Lambek calculus and references on this
topic. Our investigation into the role of ternary relations in
convolution was sparked by an insightful comment from one of our
anonymous reviewers for \cite{DongolHS16}.  We are also grateful to
Victor Gomes for his joint work on the Isabelle formalisation, and
Alasdair Armstrong for additional Isabelle advice.

\bibliographystyle{alpha}
\bibliography{interval}

\begin{thebibliography}{HMSW11}

\bibitem[AD93]{AllweinD93}
G.~Allwein and J.~M. Dunn.
\newblock Kripke models for linear logic.
\newblock {\em Journal of Symbolic Logic}, 58(2):514--545, 1993.

\bibitem[AFP]{AFP}
Archive of {F}ormal {P}roofs.
\newblock \url{https://www.isa-afp.org/index.shtml}.

\bibitem[AGS16]{ArmstrongGS16}
A.~Armstrong, V.~B.~F. Gomes, and G.~Struth.
\newblock Building program construction and verification tools from algebraic
  principles.
\newblock {\em Formal Aspects of Computing}, 28(2):265--293, 2016.

\bibitem[All83]{All83}
J.~F. Allen.
\newblock Maintaining knowledge about temporal intervals.
\newblock {\em Commun. ACM}, 26(11):832--843, 1983.

\bibitem[AM94]{AndrekaMikulas94}
H.~Andr{\'e}ka and S.~Mikul{\'a}s.
\newblock {L}ambek calculus and its relational semantics: Completeness and
  incompleteness.
\newblock {\em J.\ Logic, Language and Information}, 3(1):1--37, 1994.

\bibitem[AV93]{AbramskyV93}
S.~Abramsky and S.~Vickers.
\newblock Quantales, observational logic and process semantics.
\newblock {\em Mathematical Structures in Computer Science}, 3(2):161--227,
  1993.

\bibitem[BdRV01]{BlackburndRV01}
P.~Blackburn, M.~de~Rijke, and Y.~Venema.
\newblock {\em Modal Logic}.
\newblock Cambridge University Press, 2001.

\bibitem[BR84]{BerstelReutenauer}
J.~Berstel and C.~Reutenauer.
\newblock {\em Les s\'eries rationnelles et leurs langagues}.
\newblock Masson, 1984.

\bibitem[BRZ00]{BRC00}
R.~Barua, S.~Roy, and C.~Zhou.
\newblock Completeness of neighbourhood logic.
\newblock {\em J. Log. Comput.}, 10(2):271--295, 2000.

\bibitem[CDS20a]{CranchDS20}
J.~Cranch, S.~Doherty, and G.~Struth.
\newblock Convolution and concurrency.
\newblock {\em CoRR}, abs/2002.02321, 2020.

\bibitem[CDS20b]{CranchDS20b}
J.~Cranch, S.~Doherty, and G.~Struth.
\newblock Relational semigroups and object-free categories.
\newblock {\em CoRR}, abs/2001.11895, 2020.

\bibitem[CM16]{ITLHomepage}
A.~Cau and B.~Moszkowski.
\newblock {ITL} homepage.
\newblock \url{http://antonio-cau.co.uk/ITL/itlhomepage.html}, 2016.
\newblock Fetched 18-03-16.

\bibitem[Con71]{Conway71}
J.~H. Conway.
\newblock {\em Regular Algebra and Finite Machines}.
\newblock Chapman and Hall, 1971.

\bibitem[Con95]{Connes95}
A.~Connes.
\newblock {\em Noncommutative Geometry}.
\newblock Academic Press, 1995.

\bibitem[CP61]{CliffordP61}
A.~H. Clifford and G.~B. Preston.
\newblock {\em The Algebraic Theory of Semigroups I}.
\newblock American Mathematical Society, 1961.

\bibitem[DGHS17]{DongolGHS17}
B.~Dongol, V.~B.~F. Gomes, I.~J. Hayes, and G.~Struth.
\newblock Partial semigroups and convolution algebras.
\newblock {\em Archive of Formal Proofs}, 2017.
\newblock \url{https://www.isa-afp.org/entries/PSemigroupsConvolution.shtml}.

\bibitem[DGS15]{DongolGS15}
B.~Dongol, V.~B.~F. Gomes, and G.~Struth.
\newblock A program construction and verification tool for separation logic.
\newblock In {\em {MPC} 2015}, volume 9129 of {\em LNCS}, pages 137--158.
  Springer, 2015.

\bibitem[DHS16]{DongolHS16}
B.~Dongol, I.~J. Hayes, and G.~Struth.
\newblock Convolution as a unifying concept: Applications in separation logic,
  interval calculi, and concurrency.
\newblock {\em {ACM} Trans. Comput. Log.}, 17(3):15, 2016.

\bibitem[DMS06]{DesharnaisMS06}
J.~Desharnais, B.~M{\"{o}}ller, and G.~Struth.
\newblock Kleene algebra with domain.
\newblock {\em {ACM} TOCL}, 7(4):798--833, 2006.

\bibitem[Dos92]{Dosen92}
K.~Dosen.
\newblock A brief survey of frames for the {L}ambek calculus.
\newblock {\em Mathematical Logic Quarterly}, 38(1):179--187, 1992.

\bibitem[DR02]{DunnRestall02}
J.~M. Dunn and G.~Restall.
\newblock Relevance logic.
\newblock In D.~Gabbay and F.~Guenthner, editors, {\em Handbook of
  Philosophical Logic}, volume~6, pages 1--128. Kluwer, 2002.

\bibitem[FB94]{FoulisB94}
D.~J. Foulis and M.~K. Bennett.
\newblock Effect algebras and unsharp quantum logics.
\newblock {\em Foundations of Physics}, 24(10):1331--1352, 1994.

\bibitem[GL06]{GalmicheL06}
D.~Galmiche and D.~Larchey{-}Wendling.
\newblock Expressivity properties of boolean {BI} through relational models.
\newblock In {\em {FSTTCS} 2006}, volume 4337 of {\em LNCS}, pages 357--368.
  Springer, 2006.

\bibitem[GMS04]{GorankoMS04}
V.~Goranko, A.~Montanari, and G.~Sciavicco.
\newblock A road map of interval temporal logics and duration calculi.
\newblock {\em Journal of Applied Non-Classical Logics}, 14(1-2):9--54, 2004.

\bibitem[Gog67]{Goguen67}
J.~A. Goguen.
\newblock L-fuzzy sets.
\newblock {\em J. Mathematical Analysis and Applications}, 18:145--174, 1967.

\bibitem[GS16]{GomesS16}
V.~B.~F. Gomes and G.~Struth.
\newblock Modal {K}leene algebra applied to program correctness.
\newblock In {\em {FM} 2016}, volume 9995 of {\em LNCS}, pages 310--325, 2016.

\bibitem[Hei25]{Heisenberg25}
W.~Heisenberg.
\newblock {\"U}ber quantentheoretische {U}mdeutung kinematischer und
  mechanischer {B}eziehungen.
\newblock {\em Z. Physik}, 33:879--893, 1925.

\bibitem[HM08]{HM08}
P.~H{\"o}fner and B.~M{\"o}ller.
\newblock Algebraic neighbourhood logic.
\newblock {\em J. Log. Algebr. Program.}, 76(1):35--59, 2008.

\bibitem[HM09]{HM09}
P.~H{\"o}fner and B.~M{\"o}ller.
\newblock An algebra of hybrid systems.
\newblock {\em J. Log. Algebr. Program.}, 78(2):74--97, 2009.

\bibitem[HMSW11]{HMSW11}
T.~Hoare, B.~M{\"o}ller, G.~Struth, and I.~Wehrman.
\newblock Concurrent {Kleene} algebra and its foundations.
\newblock {\em J. Log. Algebr. Program.}, 80(6):266--296, 2011.

\bibitem[HS91]{HS91}
J.~Y. Halpern and Y.~Shoham.
\newblock A propositional modal logic of time intervals.
\newblock {\em J. ACM}, 38(4):935--962, October 1991.

\bibitem[HWW18]{HardingWW18}
J.~Harding, C.~Walker, and E.~Walker.
\newblock The convolution algebra.
\newblock {\em Algebra Universalis}, 79:1--33, 2018.

\bibitem[JT51]{JonssonT51}
B.~J\'onsson and A.~Tarski.
\newblock Boolean algebras with operators. {Part {I}}.
\newblock {\em American Journal of Mathematics}, 73(4):891--939, 1951.

\bibitem[Kon13]{Konur}
S.~Konur.
\newblock A survey on temporal logics for specifying and verifying real-time
  systems.
\newblock {\em Frontiers of Computer Science}, 7(3):370--403, 2013.

\bibitem[Lam58]{Lambek58}
J.~Lambek.
\newblock The mathematics of sentence structure.
\newblock {\em Americal Mathematical Monthly}, 65:154--170, 1958.

\bibitem[Mac98]{MacCaull98}
W.~MacCaull.
\newblock Relational semantics and a relational proof-calculus for full
  {L}ambek calculus.
\newblock {\em J.\ Symbolic Logic}, 63(2):623--637, 1998.

\bibitem[MGMS11]{MonicaGMS11}
D.~Della Monica, V.~Goranko, A.~Montanari, and G.~Sciavicco.
\newblock Interval temporal logics: a journey.
\newblock {\em Bulletin of the {EATCS}}, 105:73--99, 2011.

\bibitem[ML98]{MacLane98}
S.~Mac~Lane.
\newblock {\em Categories for the Working Mathematician}.
\newblock Springer, second edition, 1998.

\bibitem[MM83]{MM83}
B.~C. Moszkowski and Z.~Manna.
\newblock Reasoning in interval temporal logic.
\newblock In {\em Logic of Programs}, volume 164 of {\em LNCS}, pages 371--382.
  Springer, 1983.

\bibitem[MM11]{MM11}
J.~Marcinkowski and J.~Michaliszyn.
\newblock The ultimate undecidability result for the {Halpern-Shoham} logic.
\newblock In {\em {LICS}}, pages 377--386. {IEEE} Computer Society, 2011.

\bibitem[Mos12]{Mos2012-LMCS}
B.~C. Moszkowski.
\newblock A complete axiom system for propositional interval temporal logic
  with infinite time.
\newblock {\em Logical Methods in Computer Science}, 8(3), 2012.

\bibitem[MR12]{MootR12}
R.~Moot and C.~Retor\'e.
\newblock {\em The Logic of Categorial Grammars}, volume 6850 of {\em LNCS}.
\newblock Springer, 2012.

\bibitem[PR98]{PandyaR98}
P.~K. Pandya and Y.~S. Ramakrishna.
\newblock Recursive mean-value calculus.
\newblock In {\em {FSTTCS}}, volume 1530 of {\em LNCS}, pages 257--268.
  Springer, 1998.

\bibitem[PRS09]{PaunRS09}
G.~Paun, G.~Rozenberg, and A.~Salomaa, editors.
\newblock {\em The Oxford Handbook of Membrane Computing}.
\newblock Oxford University Press, 2009.

\bibitem[Rey02]{Reynolds02}
J.~C. Reynolds.
\newblock Separation logic: {A} logic for shared mutable data structures.
\newblock In {\em LICS 2002}, pages 55--74. {IEEE} Computer Society, 2002.

\bibitem[Ros90]{Rosenthal90}
K.~L. Rosenthal.
\newblock {\em Quantales and Their Applications}.
\newblock Longman Scientific $\&$ Technical, 1990.

\bibitem[Ros91]{Rosenthal91}
K.~L. Rosenthal.
\newblock Free quantaloids.
\newblock {\em J.\ Pure and Applied Algebra}, 72:67--82, 1991.

\bibitem[Ros97]{Rosenthal97}
K.~L. Rosenthal.
\newblock Relational monoids, multirelations, and quantalic recognizers.
\newblock {\em Cahiers de Topologie et G\'eom\'etrie Diff\'erentielle
  Cat\'egoriques}, 38(2):161--171, 1997.

\bibitem[Rot64]{Rota64}
G.-C. Rota.
\newblock On the foundations of combinatorial theory {I}: Theory of {M}\"obius
  functions.
\newblock {\em Zeitschrift f\"ur Wahrscheinlichkeitstheorie und verwandte
  Gebiete}, 2(4):340--368, 1964.

\bibitem[Sko20]{Skolem20}
T.~Skolem.
\newblock Logisch-kombinatorische {U}ntersuchungen \"uber die {E}rf\"ullbarkeit
  oder {B}eweisbarkeit mathematischer {S}\"atze nebst einem {T}heorem \"uber
  dichte {M}engen.
\newblock {\em Videnskabsakademiet i Kristiania}, 1(4):1--36, 1920.

\bibitem[Ven90]{Venema90}
Y.~Venema.
\newblock Expressiveness and completeness of an interval tense logic.
\newblock {\em Notre Dame Journal of Formal Logic}, 31(4):529--547, 1990.

\bibitem[Ven91]{Venema91}
Y.~Venema.
\newblock A modal logic for chopping intervals.
\newblock {\em Journal of Logic and Computation}, 1(4):453--476, 1991.

\bibitem[ZH04]{ZH04}
C.~Zhou and M.~R. Hansen.
\newblock {\em Duration Calculus: A Formal Approach to Real-Time Systems}.
\newblock EATCS: Monographs in Theoretical Computer Science. Springer, 2004.

\bibitem[ZHL95]{ChaochenHX95}
C.~Zhou, D.~V. Hung, and X.~Li.
\newblock A duration calculus with infinite intervals.
\newblock In {\em {FCT}}, volume 965 of {\em LNCS}, pages 16--41. Springer,
  1995.

\end{thebibliography}


\newpage
\appendix

\newcommand\Tstrut{\rule{0pt}{2.6ex}}         
\newcommand\Bstrut{\rule[-0.9ex]{0pt}{0pt}}   

\section{Glossary of Algebraic Structures}\label{S:alg-summary}


\small 
Within the table the number of each definition is given.

\centering
\begin{tabular}[t]{|l|l|}
  \hline
  Structure  &Axioms\\
  \hline
  \hline \Tstrut
  \ref{def:partial-semigroup} partial semigroup  
             & $X$ is a set \\
  $(X,\cdot,D)$ 
             & $D\subseteq X \times X$ is the domain of
               definition of composition \\
             & $\cdot: D\to X$ is an associative operation, i.e.,  \\
             & \qquad $D\,  x\,  y\, \wedge\,  D\, (x \cdot y)\, z \Leftrightarrow 
               D\, y\, z\,  \wedge  D\, x\,  (y \cdot  z)$ \\  
             & \qquad $D\, x\, y\, \wedge\, D\,  (x \cdot y)\,  z
               \Rightarrow (x \cdot y) \cdot z = x \cdot (y \cdot z)$  
               \Bstrut
  \\\hline \Tstrut
  \ref{def:partial-monoid} partial monoid
             & $(X, \cdot, D)$ is a partial semigroup \\
  $(X,\cdot,D,E)$ 
             & $E\subseteq X$ is a set of (generalised) units, i.e., \\
             & \qquad $\exists e \in E.\  D\,  e\,  x  \,\wedge\,  e \cdot x = x$ \\
             & \qquad$\exists e \in E.\  D\,  x\,  e  \,\wedge\,  x \cdot e = x$ \\ 
             & \qquad$e_1,e_2 \in E \,\wedge\,  D\,  e_1\,  e_2\,\,\Rightarrow\,\, e_1 = e_2$
               \Bstrut
  \\\hline \Tstrut
  \ref{def:quantale} quantale
             & $(X,\le)$ is a complete lattice with least and greatest elements $0$ and $\top$, \\
  $(X,\le,\cdot)$ & $(X,\cdot)$ is a semigroup, composition is
                    left and right distributive, i.e., \\
             & $x \cdot \Sup\, Y =  \Supr{y\in Y}{}{x \cdot y}$ and 
               $(\Sup\, X)\cdot y = \Supr{x\in X}{}{x \cdot y}$
               \Bstrut
  \\\hline \Tstrut
  \ref{def:quantale} unital quantale 
             & quantale such that $(X,\cdot,1)$ is a monoid with unit $1$ 
               \Bstrut 
  \\\hline \Tstrut
  \ref{def:quantale} distrib. quantale 
             & quantale with $x\sqcap (\Sup Y) = \Sup_{y \in Y} (x\sqcap y)$ and 
               $x \sqcup (\Inf Y) = \Inf_{y\in Y} (x \sqcup y)$ 
               \Bstrut 
  \\\hline \Tstrut
  \ref{def:quantale} boolean quantale 
             & distributive quantale with   $x\sqcap \overline{x} = 0$ and $x\sqcup \overline{x} = \top$ 
               \Bstrut 
  \\\hline \Tstrut
  \ref{def:weak-quantale} weak quantale 
             & quantale with left distributivity  weakened to \\
             &  $Y\neq\emptyset  \imp x\cdot \bigsqcup Y= \bigsqcup_{y\in Y} x\cdot y$ 
               \Bstrut 
  \\\hline \Tstrut
  \ref{def:proto-quantale} proto-quantale 
             & quantale with possibly non-associative multiplication 
               \Bstrut 
  \\\hline \Tstrut
  \ref{def:relational-semigroup} relational 
             & $X$ is a set, \\
  \quad\quad\quad semigroup
              & $R \subseteq X \times X \times X$ satisfies relational associativity, i.e., \\
  $(X,R)$ 
            & \qquad $  (\exists y\in X.\ R\, y\, u\, v\wedge R\, x\, y\, w) \Leftrightarrow 
               (\exists z\in X.\ R\,  z\, v\, w\wedge R\, x\, u\, z)$ 
               \Bstrut 
  \\\hline \Tstrut
  \ref{def:relational-monoid} relational monoid 
             & $(X, R)$ is a relational semigroup, \\
  $(X,R,\xi)$ 
             &  $\xi\subseteq X$ a set of units, i.e.,\\
             & \qquad $\exists e\in \xi.\ R\, x\, e\, x$ \\
             & \qquad $\exists e\in \xi.\ R\, x\, x\, e$ \\
             & \qquad $e\in \xi\wedge  R\, x\, e\, y  \Rightarrow x=y$\\
             & \qquad $e\in \xi\wedge  R\, x\, y\, e  \Rightarrow x=y$
               \Bstrut 
  \\\hline \Tstrut
  \ref{def:semidirect-product} semidirect  
             & $(S,\odot, D_\odot)$ and $(T,\oplus,D_\oplus)$ are partial semigroups \\
  \quad\quad\quad product
             & $D_{\sddot} = \{((s_1,t_1),(s_2,t_2)) \mid D_\odot\, s_1\, s_2\, \wedge\,
               D_\circ\, s_1\, t_2\, \wedge\, D_\oplus\, t_1\, (s_1\circ t_2)\}$, where \\
  $(S\times T,\sddot,D_{\sddot})$ 
             & \quad $\circ: D_\circ \to T$ such that $D_\circ \subseteq S\times T$ satisfies \\
             & \quad $D_\odot\, s_1\, s_2\, \wedge\,  D_\circ\,  (s_1 \odot s_2)\,  t \Leftrightarrow
               D_\circ\, s_2\, t\, \wedge\,  D_\circ\,  s_1\,  (s_2 \circ t)$ \\
             & \quad $D_\odot\, s_1\, s_2\, \wedge\,  D_\circ\, (s_1
               \odot s_2)\, t  \Rightarrow   s_1 \circ (s_2  \circ t)
               = (s_1 \odot s_2) \circ t$ \\  
             & \quad $D_\oplus\, t_1\, t_2\, \wedge\,  D_\circ\, s_1\,
               (t_1 \oplus t_2) \Leftrightarrow  D_\circ\, 
               s_1\, t_1\, \wedge\,  D_\circ\, s_1\, t_2\,  \wedge\,
               D_\oplus\, (s_1 \circ t_1)\, 
               (s_1 \circ t_2)$ \\ 
             & \quad $D_\oplus\, t_1\, t_2\, \wedge\, D_\circ\, s_1\,
               (t_1 \oplus t_2) \Rightarrow  (s_1 \circ t_1) \oplus  (s_1 \circ t_2) = s_1 \circ (t_1 \oplus  t_2)$\\
             & \quad $(s_1,t_1)\sddot (s_2,t_2) =  (s_1\odot s_2,t_1 \oplus  (s_1\circ t_2))$
               \Bstrut 
  \\\hline \Tstrut
  \ref{def:quantale-module} quantale module  
             & $\calQ = (Q, \le_\calQ, \cdot)$ is a quantale, $\calL =
               (L, \preceq_\calL)$ a complete lattice\\ 
             & $\circ: \calQ \to \calL \to \calL$\\
             & for all $u,v\in Q$, $x\in L$, $V \subseteq Q$ and
               $X \subseteq L$, \\
             & \qquad 
               $(u \cdot v)\circ x
               = u \circ (v \circ x)$ \\
             & \qquad $(\Sup V) \circ x 
               = \Supr{v \in V}{}{v \circ x}$ \\
             & \qquad $u \circ \Sup X 
               = \Supr{x \in X}{}{u \circ x}$ \\
             & \qquad $1\circ x=x$ holds whenever $\calQ$ is unital 
  \\\hline
\end{tabular}


\section{Cross-References to Archive of Formal Proofs}\label{S:crossref}

\small
\begin{tabular}[t]{ll}
  Result in article & Result  in Archive of Formal Proofs  \\
\hline
  \hline \Tstrut
  \reflem{P:unary-module} 
          &    {\tt fdia-Un-rel}, {\tt fdia-Sup-fun}, {\tt fdia-seq}, {\tt fdia-Id}
  \\[1mm]
    \reflem{P:unary-mod-galois} 
          & {\tt fdia-bbox-galois},
           {\tt bdia-fbox-galois} 
  \\[1mm]
  \reflem{P:unary-mod-conjugation} 
          & {\tt dia-conjugate}, {\tt box-conjugate} 
  \\[1mm]
  \reflem{P:binary-module} & {\tt bmod-Un-rel}, {\tt bmod-Sup-fun1}, {\tt bmod-Sup-fun2} \\[1mm]
  \reflem{P:umod-bmod} 
          & {\tt fdia-bmod-comp},
          {\tt bdia-bmod-comp} \\[1mm]
  \reflem{P:bmod-umod} 
          & {\tt bmod-fdia-comp},
           {\tt bmod-fdia-comp-var} \\[1mm]
 \reflem{P:assoccounter} & Nitpick finds counterexample (b) after
                           Lemma {\tt rel-fun-assoc} \\[1mm]
  \reflem{P:bimod-assoc} 
          & \tt rel-fun-assoc\\[1mm]
  \reflem{P:unitcounter} & Nitpick finds (different) counterexample
                            after Lemma\\
& {\tt rel-fun-assoc-weak} \\[1mm]
  \reflem{P:bimod-unit} 
          & unnamed lemmas after  \tt bmod-oner\\[1mm]
  \refthm{P:bmod-lifting} 
          & second interpretation {\tt rel-fun},\\
& first and fourth interpretation {\tt rel-fun2}\\[1mm]
  \refprop{P:res-lifting} 
          & {\tt bmod-comp-bres-galois}, {\tt bmod-comp-fres-galois}\\[1mm]
  \reflem{P:ps-rs} 
          & sublocales {\tt rel-partial-semigroup},  {\tt rel-partial-monoid} \\[1mm]
  \refcor{P:old-lifting} 
          & first and second instantiation {\tt fun} in Sec 7.2 \\[1mm]
Example \ref{ex:ord-pairs} & instatiation \tt dprod\\[1mm]
Example \ref{ex:part-prod-monoid} & instantiations \tt prod\\[1mm]
Example \ref{ex:monoid-set-monoid} & interpretations {\tt ps-prod},
                                     {\tt pm-prod}\\[1mm]
  \reflem{P:seg-sg} 
          & instantiations {\tt dprod}, 
           {\tt segment} \\[1mm]
     \refcor{P:seg-lifting} & via instantiations \reflem{P:seg-sg}, \refcor{P:old-lifting}\\[1mm]
  \refprop{P:semigroup-sdprod} 
          & sublocale statement {\tt dp-semigroup} \\[1mm]
  \refprop{P:wquantale-sdprod} 
          & interpretations {\tt dp-quantale} and {\tt dpu-quantale}\\[1mm]
  \refthm{P:weak-quantale-lifting} 
          & second and third interpretation {\tt rel-fun}\\
& fourth interpretation {\tt rel-fun}\\[1mm]
  \refprop{P:wq-seg-lifting} &covered by
                               \refprop{P:semigroup-sdprod},
                               \refprop{P:wquantale-sdprod},
                               \refthm{P:weak-quantale-lifting} \\
& except right unit law and (d)
\\[1mm]
 \refprop{P:itl} & covered by Example~\ref{ex:monoid-set-monoid},
                   \refcor{P:old-lifting},
                   \refprop{P:semigroup-sdprod},
                   \refprop{P:wq-seg-lifting}
\end{tabular}

\bigskip 

\justify


Results not formalised with Isabelle:
\refcor{P:locfin-lifting}, 
\reflem{P:rel_embedding},
\reflem{P:open-seg-sg},
\reflem{P:better-open-seg-sg}, 
\reflem{P:allen-rel},
\reflem{P:ven-sem},
\reflem{P:allen_ven},
\reflem{P:hs-sem},
\refprop{P:general-duration-quantale},
\refcor{P:duration-quantale},
and \refcor{lem:mvc-2}

\end{document}